\documentclass[12pt]{article}
\usepackage{amsmath}
\usepackage{graphicx}
\usepackage{natbib}
\usepackage{url} % not crucial - just used below for the URL 

\usepackage{placeins}
\usepackage{booktabs}

\usepackage{lineno}

\usepackage[title]{appendix}

\usepackage{pdfpages}
\usepackage[normalem]{ulem}
\usepackage{xcolor}

\usepackage{comment}
\usepackage{sectsty}
\usepackage{lineno}

\newtheorem{definition}{Definition}
\newtheorem{lemma}{Lemma}
\newtheorem{theorem}{Theorem}

\newcommand{\blind}{0}

\begin{document}
%\linenumbers
%\bibliographystyle{natbib}

\def\spacingset#1{\renewcommand{\baselinestretch}%
{#1}\small\normalsize} \spacingset{1}

%%%%%%%%%%%%%%%%%%%%%%%%%%%%%%%%%%%%%%%%%%%%%%%%%%%%%%%%%%%%%%%%%%%%%%%%%%%%%%

\if0\blind
{
  \title{\bf A unifying theory for the evaluation of conditional Akaike information for mixed-effects models}
  \author{Nan Zheng\\
  	Department of Mathematics and Statistics,\\ Memorial University of Newfoundland\\
  	and \\
  	Noel G. Cadigan\thanks{Email: Noel.Cadigan@mi.mun.ca
  	}\hspace{.2cm} \\
  	Centre for Fisheries Ecosystems Research,\\ Fisheries and Marine Institute of Memorial University of Newfoundland\\
  	and \\
  	James T. Thorson\\
  	Resource Ecology and Fisheries
  	Management,\\ Alaska Fisheries Science
  	Center,\\ National Marine Fisheries Service,\\
  	National Oceanic and Atmospheric
  	Administration
  }
  \maketitle
} \fi

\if1\blind
{
  \bigskip
  \bigskip
  \bigskip
  \begin{center}
    {\LARGE\bf A unifying theory for the evaluation of conditional Akaike information for mixed-effects models}
\end{center}
  \medskip
} \fi

\bigskip
\begin{abstract}
	\{Simulation studies are conducted to assess the accuracy of our method for evaluating the conditional AIC and its model selection performance across a wide range of mixed-effects models. The proposed approach provides a valuable tool for model selection in the context of mixed-effects modeling.\}\\
	We propose two methods to evaluate the conditional Akaike information (cAI) for mixed-effects models with no restriction on cluster size. Method 1 is designed for continuous data and includes formulae for the derivatives of fixed and random effects estimators with respect to observations. Method 2, compatible with any type of observation, requires modeling the marginal (or prior) distribution of random effects as a multivariate normal distribution. Simulations show that Method 1 performs well with Gaussian data but struggles with skewed continuous distributions, whereas Method 2 consistently performs well across various distributions, including normal, gamma, negative binomial, and Tweedie, with flexible link functions. A case study demonstrates the differences in model selection for real-world data between the conventional AIC and the conditional AIC. Based on our findings, we recommend Method 2 as a distributionally robust cAI criterion for model selection in mixed-effects models.	
%The text of your abstract.  200 or fewer words.
\end{abstract}

\noindent%
{\it Keywords:}  Mixed-effects model; Model selection; Marginal likelihood; Conditional likelihood; Maximum marginal likelihood estimator; Empirical Bayes estimator  %3 to 6 keywords, that do not appear in the title
\vfill

\newpage
\spacingset{1.75} % DON'T change the spacing!
\section{Introduction}
\label{sec:intro}
Mixed-effects models (MMs) are widely used across disciplines such as medicine, public health, pharmacology, fisheries, and ecology, where the quantities of interest are linear or nonlinear functions of both fixed and random effects \citep{lindstrom1990nonlinear}. In these models, the log-joint likelihood of the data ($y$) and random effects (REs, $\Psi$) conditional on the fixed effects (parameters, $\theta$) can be expressed as $\log{ f(y,\Psi|\theta)} = \log{ f(y|\theta,\Psi)} + \log{ f(\Psi|\theta)}$. Maximum likelihood estimation of $\theta$ is achieved by optimizing the log-marginal likelihood, $\log{ f(y|\theta)}$, which results from integrating out the REs from the joint density. 

Practitioners of MMs often need to select the optimal model from a wide range of alternatives \citep[e.g.,][]{cadigan2022complex}, with many relying on the conventional Akaike information criterion \citep[AIC,][]{Akaike1973AIC}. AIC is intended to estimate the expected log-marginal likelihood for out-of-sample predictions, where the degree of freedom correction adjusts for the expected difference between in-sample and out-of-sample prediction \citep[termed ``model optimism" in][]{hooten2015guide}. However, \cite{vaida2005conditional} pointed out that when the REs are of primary interest, rather than merely a device for modeling the correlation among responses, the Akaike Information (AI) and the corresponding model selection criterion should be redefined using the log-conditional likelihood, $\log{ f(y|\hat{\theta}, \hat{\Psi})}$. These are termed conditional AI (cAI) and conditional AIC (cAIC), respectively. Here and throughout the paper, the hat notation denotes estimators of parameters or predictors of REs. Studies also suggest that the conventional AIC is too conservative for mixed-effects model selection, while the cAIC performs more reliably in this context \citep[e.g.][]{yu2012conditional,wood2016smoothing}.

Given the strong relevance of cAIC to mixed-effects model applications, its formulation and evaluation have been extensively studied across various settings, including linear mixed-effects models with independent Gaussian observations of known variance \citep{liang2008note}, generalized linear mixed models with known dispersion \citep{yu2012conditional}, and smooth models \citep{wood2016smoothing}, as well as broader efforts aimed at developing a unified framework for cAIC \citep[e.g.,][]{donohue2011conditional,saefken2014unifying}. Despite this substantial body of work, a widely validated unifying methodology for cAIC remains lacking. For example, the approach of \citet{wood2016smoothing} is restricted to smooth models; \citet{saefken2014unifying} demonstrated that the method of \citet{donohue2011conditional} performs poorly and that of \citet{yu2012conditional} exhibits a tendency toward underestimation in their simulation studies. Moreover, although \citet{saefken2014unifying} and \citet{safken2021conditional} derive exact expressions for cAIC using techniques developed by \citet{stein1972bound} and \citet{hudson1978natural}, their methodology is currently limited to Gaussian, exponential, and Poisson response distributions. 

This lack of a unified cAIC evaluation methodology makes it difficult for practitioners of MMs to identify if there are appropriate cAIC formulae in the literature that correspond to their specific research settings. This challenge is further compounded by the development of advanced modeling tools like TMB \citep{kristensen2016tmb}, which enable the exploration of MMs that do not easily fit into the existing cAIC frameworks. On the other hand, some existing cAIC methods may be further generalized and improved using these advanced modeling tools. For instance, \cite{liang2008note} suggests evaluating their cAIC using numerical approximations based on small perturbations of the observed data, which can be challenging to implement in practice. Leveraging R \citep{r2018citation} and TMB's strong automatic differentiation (up to order three) capabilities, we propose a simpler method in Sec. \ref{sec:meth}, applicable to broader scenarios including correlated observations, not addressed by \cite{liang2008note}. This example also highlights our aim: to explore cAIC evaluation using recent advances in numerical methods and corresponding inferential theory for MMs \citep[e.g.,][]{zheng2021frequentist,Zheng2021P2}.

The rest of the paper is organized as follows. In Sec.~\ref{sec:meth}, we introduce two cAIC formulas. The first formula, suited for continuous observations, requires evaluation using provided derivatives of the parameter and RE estimators with respect to the data. The second formula, which accommodates all observation types, assumes that the REs are modeled with a multivariate normal distribution. In Sec.~\ref{sec:simulation_October_21_2024}, we perform extensive simulations to test both cAIC formulas across different data types and RE structures. A fish stock assessment model selection example in Sec.~\ref{sec:fish_stock_model_selection} illustrates the application of the proposed cAIC method. Concluding remarks are presented in Sec.~\ref{sec:conc}.

\section{Methods}
\label{sec:meth}
\subsection{Notation and conditional AIC}\label{notation_October_8_2024}
Let $y, \theta$ and $\Psi$ represent the vectors of all observations, assumed fixed effects (parameters) and assumed REs, respectively. The log-joint likelihood is denoted as
\begin{linenomath*} 
	\begin{align}\label{EQ:log_joint_likelihood_October_8_2024}
		\begin{split}
			l_j(y,\theta,\Psi) &= \log{ f(y,\Psi|\theta)} = \log{ f(y|\theta,\Psi)} + \log{ f(\Psi|\theta)} = l_c(\theta,\Psi|y) + l_r(\theta,\Psi).
		\end{split}	 
	\end{align}
\end{linenomath*} 
Assume that the true distribution of $y$ is $g(y|u)$, where $u$ is the true REs vector with distribution $p(u)$. The vector $u$ may differ in length from $\Psi$, but we assume that $g(y|u)=f(y|\theta_o,\Psi_o)$ for some $\Psi_o$ and $\theta_o$. This assumption is a common requirement, explicit or implicit, in various derivations of cAIC \citep[see, e.g.,][]{vaida2005conditional,liang2008note}, although some studies have addressed cases where the candidate model does not contain the true model in linear mixed models \citep[e.g.,][]{kawakubo2014modified}. Notably, $\theta_o$ and $\Psi_o$ may have different dimensions from $\theta$ and $\Psi$, respectively. For instance, if $u$ is longer than $\Psi$, then $\Psi_o$ may match the length of $u$, while $\Psi$ assumes that some components of $\Psi_o$ are identical. Let $\hat{\theta}$ denote the maximum marginal likelihood estimator (MMLE) of $\theta$, and $\hat{\Psi}(y)$ the empirical Bayes estimator of $\Psi$ \citep{kass1989approximate}. 

\cite{vaida2005conditional} defined the cAI as follows.
\begin{definition}
	The conditional Akaike information is defined to be
	\begin{linenomath*} 
		\begin{align}\label{EQ:cAI_October_9_2024}
			\begin{split}
				\mathrm{cAI} &= -2 \,\mathrm{E}_{g(y,u)}\left( \mathrm{E}_{g(y^*|u)}\left[ \log f\{ y^*|\hat{\theta}(y), \hat{\Psi}(y) \} \right] \right) \\
				&= -2\int \log f\{ y^*|\hat{\theta}(y), \hat{\Psi}(y) \}\, g(y^*|u) g(y|u)p(u) dy^*dydu,
			\end{split}	 
		\end{align}
	\end{linenomath*}
	where $y^*$ is the prediction dataset which is independent of $y$ conditional on $u$, and from the same distribution $g(\cdot|u)$ as $y$.
\end{definition}
Given the assumption $g(y|u) = f(y|\theta_o, \Psi_o)$, the conditional expectations with respect to $y^*$ and $y$ can be computed. However, since $p(u)$ is unknown, the derivation of cAIC should avoid evaluating the expectation over $u$.

Following \cite{vaida2005conditional} and \cite{liang2008note}, we define a cluster as a set of observations that share the same RE component. Let $T$ denote the total number of clusters, and $n_t$ the number of observations in the $t^{\mathrm{th}}$ cluster.

\subsection{Evaluation of conditional AIC}\label{evaluation_cAIC_October_9_2024}
The cAIC formula presented in \cite{liang2008note} includes the derivative of the mean estimators of $y$ with respect to $y$, which can be expressed as the derivatives of $\hat{\theta}(y)$ and $\hat{\Psi}(y)$ with respect to $y$. These derivatives are provided in the following lemma.
\begin{lemma}\label{lemma:derivatives_hattheta_Psi_wrt_y_October_10_2024}
	Assume that the log-marginal likelihood $l(y,\theta)$ and log-joint likelihood $l_j(y,\theta,\Psi)$ have well-defined second-order derivatives. For the maximum marginal likelihood estimator $\hat{\theta}(y)$, obtained by maximizing $l(y,\theta)$ with respect to $\theta$, and the empirical Bayes estimator $\hat{\Psi}(y)$, obtained by maximizing $l_j(y,\hat{\theta},\Psi)$ with respect to $\Psi$, the first-order derivatives with respect to $y$ are given by
	\begin{linenomath*} 
		\begin{align}\label{EQ:derivatives_hattheta_Psi_wrt_y_October_10_2024}
			\begin{split}
				\dfrac{d\hat{\theta}^{\top}(y)}{d y} &= -\dfrac{\partial^2 l(y,\hat{\theta})}{\partial y\partial\hat{\theta}^{\top}}\left\lbrace \dfrac{\partial^2 l(y,\hat{\theta})}{\partial\hat{\theta}\partial\hat{\theta}^{\top}} \right\rbrace^{-1},\\
				\dfrac{d\hat{\Psi}^{\top}(y)}{d y} &= -\left\lbrace \dfrac{\partial^2 l_j(y,\hat{\theta},\hat{\Psi})}{\partial y\partial\hat{\Psi}^{\top}} + \dfrac{d\hat{\theta}^{\top}(y)}{d y}\dfrac{\partial^2 l_j(y,\hat{\theta},\hat{\Psi})}{\partial \hat{\theta}\partial\hat{\Psi}^{\top}} \right\rbrace \left\lbrace \dfrac{\partial^2 l_j(y,\hat{\theta},\hat{\Psi})}{\partial \hat{\Psi}\partial\hat{\Psi}^{\top}} \right\rbrace^{-1}.
			\end{split}	 
		\end{align}
	\end{linenomath*} 
\end{lemma}
Lemma \ref{lemma:derivatives_hattheta_Psi_wrt_y_October_10_2024} follows from the vanishing of the derivatives of the log-marginal and log-joint likelihoods with respect to $\hat{\theta}$ and $\hat{\Psi}$, respectively, and an application of the chain rule (see Appendix \ref{app:derivatives_derivation_October_30_2024}). These derivatives can be efficiently computed using TMB and the ``numericDeriv()" function in R, leveraging their strong numerical differentiation capabilities (R and TMB Codes are provided in the Supplementary Material). For instance, by treating $y$ as parameters rather than data, TMB can quickly compute $\partial^2 l_j(y,\hat{\theta},\hat{\Psi})/\partial y\partial\hat{\Psi}^{\top}$ using automatic differentiation. 

Let $\Theta=(\theta^{\top},\Psi^{\top})^{\top}$. These derivatives facilitate a first-order Taylor expansion of the estimator $\hat{\Theta}(y)$ with respect to $y$. We first perform a Taylor expansion of the log-likelihood with respect to $\hat{\Theta}$, followed by a Taylor expansion of the resulting first-order term with respect to $y$. Evaluating the expectation of this expansion yields the following theorem for computing cAIC for continuous data (see Appendix \ref{app:derivation_4_October_30_2024}). Let $\mathrm{COV}_{g(y|u)}(y)$ denote the conditional covariance of $y$ given the true REs $u$.
\begin{theorem}\label{theorem_cAIC_continous_October_15_2024}
	Assume that the data $y$ follow the true density $g(y|u)=f(y|\theta_o,\Psi_o)$ for some $\theta_o$ and some REs $\Psi_o$ and $u$ with distribution $p(u)$. Let the data be modeled by a continuous density denoted by $f(y|\theta,\Psi)$. Then to the first order of Taylor expansion, cAI in (\ref{EQ:cAI_October_9_2024}) can be estimated by
	\begin{linenomath*} 
		\begin{align}\label{EQ:CAIC_October_15_2024}
			\begin{split}
				\mathrm{cAIC} = &-2\, \log f\{y|\hat{\theta}(y),\hat{\Psi}(y)\} \\
				&+ 2\, \mathrm{trace}\left\lbrace \dfrac{\partial\hat{\Theta}^{\top}(y)}{\partial y}\left. \dfrac{\partial^2 \log f(y|\Theta_o)}{\partial\Theta_o\partial y^{\top}}\right|_{y=\mu_y}\mathrm{COV}_{g(y|u)}(y) \right\rbrace,
			\end{split}
		\end{align}
	\end{linenomath*} 
	where $\mu_y=\mathrm{E}_{g(y|u)}(y)$.
\end{theorem}
Note that the modeling distribution of $\Psi$ is not required for Theorem \ref{theorem_cAIC_continous_October_15_2024}. \citet{overholser2014effective} examined cAIC for linear mixed-effects models with correlated errors, but their formulation depends on the estimated RE covariance. 

When the data are independent, normally distributed with known variance, with an identity link function, Eq. (\ref{EQ:CAIC_October_15_2024}) is exact and matches Eq. (4) in \cite{liang2008note}. Eq. (\ref{EQ:CAIC_October_15_2024}) also applies to both Gaussian and non-Gaussian continuous data with unknown correlation structures, and the second-order derivative of $\log f(y|\Theta_o)$ can be efficiently evaluated using the same method as for Lemma \ref{lemma:derivatives_hattheta_Psi_wrt_y_October_10_2024}. However, the approximation order depends on several factors, such as the link function, as well as the skewness and kurtosis of the observational distribution, and thus cannot be universally defined. In Sec.~\ref{sec:gamma_October_21_2024}, we will show that for skewed distributions, such as the gamma distribution, and when a non-identity link function is used, the performance of Eq.~(\ref{EQ:CAIC_October_15_2024}) may degrade. Additionally, we aim to accommodate discrete distributions. We therefore pursue an alternative approach in the sequel.

Derivations of marginal and conditional AIC typically rely on a second-order Taylor expansion of the observed log-likelihood $l_c(\hat{\Theta}|y)$ around $\Theta_o$ (Appendix Eq. \ref{lc_August_4_2024}) \citep[e.g.,][]{yu2012conditional,wood2016smoothing,saefken2014unifying}, with notable exceptions based on Stein's method \citep[e.g.,][]{liang2008note,saefken2014unifying}. The latter requires a canonical exponential-family formulation and a known scaling parameter and is further restricted to a limited class of observation distributions and link functions. Our unpublished numerical study suggests that, in Eq.~(\ref{lc_August_4_2024}), the linear term dominates, while the second-order term is negligible for the purpose of constructing cAIC. Consequently, our evaluation of cAIC focuses on the accurate assessment of the first-order term in Eq.~(\ref{lc_August_4_2024}). In this respect, \cite{donohue2011conditional}, \citet{yu2012conditional}, \citet{wood2016smoothing}, and \citet{saefken2014unifying} all follow the approach originally proposed by \citet{Akaike1973AIC} for deriving marginal AIC, which involves an additional Taylor expansion of $\partial l_c(\Theta_o | y)/\partial \Theta_o$ in the linear term around $\hat{\Theta}$. While this approach is appropriate in the marginal setting considered by \citet{Akaike1973AIC}, where the score function vanishes at the maximum likelihood estimator (MLE) and the MLE is asymptotically unbiased and efficient, it is less suitable in the conditional setting. Specifically, for cAIC, $\partial l_c(\hat{\Theta} | y)/\partial \hat{\Theta} \neq 0$, and $\hat{\Theta}$ is asymptotically biased when the cluster sizes $n_t$ have an upper bound \citep[e.g.,][]{Zheng2021P2}, implying that the additional Taylor expansion may introduce bias into the evaluation of cAIC. 

To circumvent this issue, we avoid further Taylor expansion of the first-order term in Eq.~(\ref{lc_August_4_2024}). Instead, as detailed in Appendix~\ref{app:derivation_6_October_30_2024}, we directly evaluate the expectation in the cAI definition (\ref{EQ:cAI_October_9_2024}) for each component of $\hat{\Theta}-\Theta_o$ appearing in the first-order term and apply integration by parts, yielding the following expressions for cAIC:
\begin{linenomath*} 
	\begin{align}\label{EQ:CAIC_exact_October_17_2024}
		\begin{split}
			\mathrm{cAIC} 	= &-2\, \log f\{y|\hat{\theta}(y),\hat{\Psi}(y)\} + 2\,p_c \\
			&+ 2\, \mathrm{E}_{g(y,u)}\left[\left\lbrace \hat{\Psi}(y)-\Psi_o\right\rbrace ^{\top}\,\dfrac{\partial\log f(y|\theta_o,\Psi_o)}{\partial\Psi_o} \right]\\
			= &-2\, \log f\{y|\hat{\theta}(y),\hat{\Psi}(y)\} + 2\,p_c + 2\,q \\
			&+ 2\,\mathrm{trace}\left(  \mathrm{E}_{p(u)}\left[\dfrac{\partial}{\partial\Psi_o^{\top}}\mathrm{E}_{g(y|u)}\{\hat{\Psi}(y)-\Psi_o\} \right] \right),  
		\end{split}
	\end{align}
\end{linenomath*}
where $p_c$ is the number of parameters in the log-conditional likelihood $\log f(y|\theta,\Psi)$, and $q$ is the number of REs. 

Evaluating the last term requires a distributional assumption about the REs. We assume the RE model, $f(\Psi|\theta)$ (not the true RE distribution $p(u)$), is a multivariate normal distribution (MVN), which is commonly used for modeling REs. Let the double dots in $\ddot{l}_j^{-1}(y,\theta,\Psi)$ and $\ddot{l}_r(\theta,\Psi)$ represent second-order derivatives with respect to $\Psi$. In Appendix~\ref{app:derivation_6_October_30_2024}, we prove the following theorem for cAIC using Eqs.~(5) and (B.1) of \cite{Zheng2021P2}.
\begin{theorem}\label{theorem_cAIC_October_18_2024}
	Assume that the data $y$ follow the true density $g(y|u)=f(y|\theta_o,\Psi_o)$ for some $\theta_o$ and some REs $\Psi_o$ and $u$ with distribution $p(u)$. Let the REs be modeled as a MVN. Then, the cAIC can be estimated as:
	\begin{linenomath*} 
		\begin{align}\label{EQ:CAIC_October_18_2024}
			\begin{split}
				\mathrm{cAIC} = &-2\, \log f\{y|\hat{\theta}(y),\hat{\Psi}(y)\} + 2\,p_c + 2\,q \\
				&- 2\,\mathrm{trace}\left\{ \ddot{l}_j^{-1}(y,\hat{\theta},\hat{\Psi})\,\ddot{l}_r(\hat{\theta},\hat{\Psi}) \right\} + O(1/n_t). 
			\end{split}
		\end{align}
	\end{linenomath*} 	
\end{theorem}
The last two terms of Eq.~(\ref{EQ:CAIC_October_18_2024}), namely, the trace term and the $O(1/n_t)$ term, both arise from the trace component of Eq.~(\ref{EQ:CAIC_exact_October_17_2024}).
If the data are uninformative about $\Psi$ then the trace term will be $\approx q$ and $\mathrm{cAIC} \approx -2\, \log f\{y|\hat{\theta}(y),\hat{\Psi}(y)\} + 2\,p_c$. On the other hand, if the data are highly informative about the $\Psi$ then the trace term will be close to zero and $\mathrm{cAIC} \approx -2\, \log f\{y|\hat{\theta}(y),\hat{\Psi}(y)\} + 2\,(p_c + q)$. Appendix \ref{app:derivation_6_October_30_2024} shows that the approximation order $O(1/n_t)$ is given by $2\,\mathrm{trace}(  \mathrm{E}_{p(u)}[(\partial\hat{\Psi}/\partial\theta^{\top})\partial\mathrm{E}_{g(y|u)}\{\hat{\theta}(y)-\theta_o\}/\partial\Psi_o^{\top} ] )$. Assuming that $p(u)$ follows a MVN with covariance matrix $\Sigma_u$, the Supplementary Material further derives that $O(1/n_t) = -2\mathrm{trace}\{\Sigma_u^{-1}(\partial\hat{\Psi}/\partial\theta^{\top})\mathrm{COV}(\hat{\theta})(\partial\hat{\Psi}^{\top}/\partial\theta)\}$. This result is also obtained using Eq. (7) in \cite{Zheng2021P2}, where the MVN assumption for $p(u)$ is unnecessary, and $\Sigma_u^{-1}$ is replaced by the modeling MVN precision matrix for $\Psi$, $\Sigma_{\Psi}^{-1}$. The marginal covariance of the MMLE, $\mathrm{COV}(\hat{\theta})$, is on the order of $O\{(Tn_t)^{-1}\}$. The precision matrices $\Sigma_u^{-1}$ and $\Sigma_{\Psi}^{-1}$ are typically sparse, and hence taking the trace over the RE dimension $T$ yields a term of order $O(1/n_t)$. It approaches zero only as cluster sizes increase. 

Some derivations of unified cAIC formulas obtain an $o(1)$ approximation by assuming that the cluster sizes $n_t$ grow unbounded as the overall sample size increases \citep[e.g.,][]{donohue2011conditional,yu2012conditional}. However, in practice, it is more common for the number of clusters to grow while cluster sizes remain bounded. Therefore, we do not eliminate the $O(1/n_t)$ term by imposing unbounded growth of cluster sizes. Instead, we justify this approximation order by comparing it with the leading-order term in the degrees of freedom \citep[DF;][]{vaida2005conditional}. For the conventional AIC, the leading term in DF, namely the number of model parameters, is of order $O(1)$. Consequently, the approximation condition in \cite{Akaike1973AIC} is satisfied with an error of order $o(1)$. In contrast, in the cAIC setting, the leading term in the DF appearing in Eq. (\ref{EQ:CAIC_October_18_2024}) corresponds to one half of the terms following $-2\, \log f\{y|\hat{\theta}(y),\hat{\Psi}(y)\}$ and is of order $O(T)$. Specifically, the terms $q$ and $- \,\mathrm{trace}\{ \ddot{l}_j^{-1}(y,\hat{\theta},\hat{\Psi})\,\ddot{l}_r(\hat{\theta},\hat{\Psi}) \}$ in Eq. (\ref{EQ:CAIC_October_18_2024}) scale linearly with the number of clusters $T$, or equivalently, the sample size. Therefore, an approximation error of order $O(1)$ is appropriate in this context. 

In the next section, we use a comprehensive simulation study to demonstrate that (\ref{EQ:CAIC_October_18_2024}) provides a reasonably accurate evaluation of the cAIC.

%Despite this, (\ref{EQ:CAIC_October_18_2024}) remains useful for model selection for two key reasons. First, the conventional AIC achieves an approximation order of $o(1)$ \citep[see, e.g.,][]{cavanaugh2019akaike}, where the degrees of freedom \citep[DF,][]{vaida2005conditional}, namely, the number of model parameters, does not increase with the number of REs or clusters. However, when the researcher's focus is on clusters instead of the population, the model DF (which corresponds to half of the terms after $-2\, \log f\{y|\hat{\theta}(y),\hat{\Psi}(y)\}$ in \ref{EQ:CAIC_October_18_2024}) increases with the number of clusters. In such cases, an approximation order of $O(1)$ is reasonable, provided the $O(1)$ term is small, as differences between mixed-effects models in cAIC are dominated by terms of order $O(T)$, such as $-2\, \log f\{y|\hat{\theta}(y),\hat{\Psi}(y)\}$, $2\,q$, and $- 2\,\mathrm{trace}\{ \ddot{l}_j^{-1}(y,\hat{\theta},\hat{\Psi})\,\ddot{l}_r(\hat{\theta},\hat{\Psi}) \}$ in (\ref{EQ:CAIC_October_18_2024}), and the competing models differ significantly in $T$. Secondly, in $O(1/n_t) = -2\mathrm{trace}\{\Sigma_u^{-1}(\partial\hat{\Psi}/\partial\theta^{\top})\mathrm{COV}(\hat{\theta})(\partial\hat{\Psi}^{\top}/\partial\theta)\}$, since $\partial\hat{\Psi}/\partial\theta^{\top}$ is typically small, its square is even smaller, making $O(1/n_t)$ negligible compared to other terms that increase with $T$.

\section{Simulation study}
\label{sec:simulation_October_21_2024}
In this simulation study, we evaluate the performance of the proposed methods for computing cAIC and conducting model selection across a range of RE structures and observation types. To our knowledge, unified approaches for evaluating cAIC include those of \citet{donohue2011conditional}, \citet{yu2012conditional}, \citet{saefken2014unifying}, the \texttt{cAIC4} R package \citep{safken2021conditional}, and \citet{wood2016smoothing}. However, \citet{donohue2011conditional} exhibited substantial underestimation in the simulation results reported by \citet{saefken2014unifying}. Moreover, both \citet{saefken2014unifying} and the \texttt{cAIC4} package currently support only Gaussian, exponential, Poisson, and binomial responses for models fitted using lme4 \citep{lme4}, which does not accommodate the broader range of RE structures and observation types considered in our study, such as the Gamma, negative binomial, and Tweedie responses. Consequently, we adopt the methods of \citet{yu2012conditional} and \citet{wood2016smoothing} as the primary benchmark comparators, and provide a separate comparison with \texttt{cAIC4} in Sec.~\ref{compare_cAIC4_August_21_2026}. %We first compare the candidate methods under a range of popular observation distributions, and then assess their performance under a more complex three-dimensional spatiotemporal RE structure.

Following the approach of \cite{liang2008note}, we examine the estimates of the bias correction, BC, which is defined by $\mathrm{cAI} = \mathrm{E}_{g(y,u)}[ -2 \,\log f\{ y|\hat{\theta}(y), \hat{\Psi}(y) \} ] + \mathrm{BC}$. The procedure involves $n_{\mathrm{out}}$ (e.g., 500) outer iterations, where REs $u$ are sampled from $p(u)$, and a dataset $y$ is generated from $g(y|u)$ to compute the expectation with respect to $g(y,u)$. Within each outer iteration, $n_{\mathrm{inner}}$ (e.g., 1000) inner iterations are performed to generate $y^*$ from $g(y^*|u)$, contributing to the expectation over $y^*$ in (\ref{EQ:cAI_October_9_2024}). The true BC, averaged over the outer iterations, is denoted by $\mathrm{BC}_{\mathrm{true}}$, while the estimated BC, averaged over the outer iterations, is denoted by $\mathrm{BC}_{\mathrm{est}}$. We then assess the relative bias, $\mathrm{RB} = (\mathrm{BC}_{\mathrm{est}}-\mathrm{BC}_{\mathrm{true}})/\mathrm{BC}_{\mathrm{true}}$, for the cAIC evaluation methods. 

\citet{wood2016smoothing} derived a BC formula for generalized additive models (GAMs), and their derivation extends directly to general MMs. Let $\beta$ denote the fixed and random effects in the likelihood $f(y|\theta,\Psi)=f(y|\beta)$. Following \citet{wood2016smoothing}, $\mathrm{BC}=2\,\mathrm{trace}[\mathrm{E}\{(\hat{\beta}-\beta_o)(\hat{\beta}-\beta_o)^{\top}|\beta_o\}\mathcal{I}_{\beta}]$, where $\hat{\beta}$, $\beta_o$ and $\mathcal{I}_{\beta}$ denote the estimator, true value, and expected negative Hessian of log-likelihood $\log\{f(y|\beta)\}$, respectively. For the conditional mean squared error (MSE) $\mathrm{E}\{(\hat{\beta}-\beta_o)(\hat{\beta}-\beta_o)^{\top}|\beta_o\}$, \citet{wood2016smoothing} proposed the estimator $V_{\beta}^*+V''$, where $V_{\beta}^*$ is the commonly used marginal covariance of $\hat{\beta}$ averaged over the prior of $\Psi_o$ \citep{kass1989approximate, zheng2021frequentist}, and $V''$ accounts for uncertainty introduced by substituting parameter estimates into $V_{\beta}^*$. Our preliminary simulations suggest that including $V''$ does not improve BC estimation, and we therefore omit it here. In $V_{\beta}^*$, the covariance of the MMLEs is given by the inverse Fisher information based on the marginal likelihood, and the covariance of the RE predictors follows from both empirical Bayes \citep{kass1989approximate} and frequentist \citep{zheng2021frequentist} formulations. TMB evaluates $V_{\beta}^*$ automatically. The resulting BC estimator, obtained by extending \citet{wood2016smoothing}, is
\begin{linenomath*} 
	\begin{align}\label{EQ:BC_GAM_December_10_2025}
		\begin{split}
			\mathrm{BC}_{\mathrm{GAM}} &= 2\,\mathrm{trace}(\hat{\mathcal{I}}_{\beta}V_{\beta}^*),
		\end{split}	 
	\end{align}
\end{linenomath*}  
where $\hat{\mathcal{I}}_{\beta}$ is the observed $\mathcal{I}_{\beta}$.

\subsection{Fisheries population dynamics examples}
\label{sec:population_dynamics_December_8_2025}
We consider a simulation setup useful in fisheries population dynamics studies to estimate
change in population size over time ($t$) at different ages ($a$) in the population. The model includes fixed age effects, $q_a$, REs for years, $\Psi_{t}$, and age-year interaction REs, $\Psi_{t,a}$, resulting in the following linear predictor:
\begin{linenomath*} 
	\begin{align}\label{EQ:RE_model_October_21_2024}
		\begin{split}
			\eta_{t,a} &= q_a + \sigma\, \Psi_{t} + \delta\, \Psi_{t,a},
		\end{split}
	\end{align}
\end{linenomath*} 
where $q_a$, $\sigma$ and $\delta$ are model parameters. Each year includes 6 age groups, indexed by subscript $a$, with $q_a=(-2, -1,  0,  1,  2,  3)$. The year REs, $\Psi_{t}$, follow an AR(1) process with autocorrelation parameter $\rho=0.8$ and marginal standard deviation (sd) of 1, while the interaction REs, $\Psi_{t,a}$, are independent standard normal variables. 

When generating data, $\sigma$ is fixed at 1 to serve as the reference RE strength, while $\delta$ takes values of 0.01, 0.1, 0.4, and 0.8 to explore different strengths of the age-year interaction REs. Based on $\eta_{t,a}$, we examine the performance of the proposed cAIC estimators for Gaussian, gamma, negative binomial and Tweedie observations, applying the appropriate link functions. All model parameters and REs are estimated; no parameters were fixed which is an improvement compared to some of the cAIC literature \citep[e.g.,][]{liang2008note}. Using the model without age-year interactions, $\Psi_{t,a}$, as the null model, we evaluate the model-selection performance of the candidate cAIC methods and the conventional (marginal) AIC as the strength of the age-year interaction increases across simulation scenarios.

%We examine the estimation of the correction term for estimating cAI defined in (\ref{EQ:cAI_October_9_2024}), using $-2 \log f\{y|\hat{\theta}(y),\hat{\Psi}(y)\}$, i.e., $\mathrm{cAI} + 2 \log f\{y|\hat{\theta}(y),\hat{\Psi}(y)\}$, which corresponds to twice the EDF. There are 500 outer iterations, each containing 1000 inner iterations. In each outer iteration, a set of REs is sampled from $p(u)$, followed by generating a dataset $y$ from $g(y|u)$, which is then used to obtain the estimators $\hat{\theta}(y)$ and $\hat{\Psi}(y)$, along with the estimated bias corrections using (\ref{EQ:CAIC_October_15_2024}) and (\ref{EQ:CAIC_October_18_2024}) for continuous data, and (\ref{EQ:CAIC_October_18_2024}) for non-continuous data. Here, the bias correction in each cAIC formula is the sum of all terms except for $-2 \log f\{y|\hat{\theta}(y),\hat{\Psi}(y)\}$. These 500 outer iterations evaluate the expectation over $y$ and $u$ as defined in (\ref{EQ:cAI_October_9_2024}). Within each outer iteration, using the sampled $u$, 1000 inner iterations are performed to generate $y^*$ from $g(y^*|u)$ for computing $\mathrm{E}_{g(y^*|u)} [ \log f\{ y^*|\hat{\theta}(y), \hat{\Psi}(y) \} ]$ in (\ref{EQ:cAI_October_9_2024}). The results are then averaged over the 500 outer iterations to compute cAI
\subsubsection{Gaussian observations}
\label{sec:Gaussian_October_21_2024}
We use the identity link function and the model for the $i$th observation $y_i$ at year $t_i$ and age $a_i$ is
\begin{linenomath*} 
	\begin{align}\label{EQ:yi_October_21_2024}
		\begin{split}
			y_i &= \eta_{t_i,a_i} + \varepsilon_i,
		\end{split}
	\end{align}
\end{linenomath*}  
where given all the REs, the error term $\varepsilon_i$ is independently normally distributed with mean 0 and sd $\sigma_e = 0.25$, 0.5 or 1. The total number of observation years is $T=50$. The sample size for each year and age point, $n_{t,a}$, is 3 or 5. When applying Eq. (\ref{EQ:CAIC_October_18_2024}), we set $n_{\mathrm{out}}=1500$ and $n_{\mathrm{inner}}=20000$. We used $n_{\mathrm{out}}=500$ and $n_{\mathrm{inner}}=1000$ when evaluating Eq. (\ref{EQ:CAIC_October_15_2024}) because of the higher computational cost of evaluating the derivatives in Eq. (\ref{EQ:derivatives_hattheta_Psi_wrt_y_October_10_2024}).

Simulation results for BC evaluation are presented in Table \ref{tab:Gaussian_simulation_October_25_2024}. \citet{liang2008note} assumed known variances for independent Gaussian observations, allowing for an exact derivation of their bias correction formula. In their simulation study \citep[Table 1,][]{liang2008note}, the highest RBs reach values of 0.024 and 0.035. Thus, if our RBs are at or below 0.035, we can reasonably consider them small. The RBs obtained from both (\ref{EQ:CAIC_October_15_2024}) and (\ref{EQ:CAIC_October_18_2024}) for estimating cAI are indeed small, with (\ref{EQ:CAIC_October_15_2024}) generally yielding slightly smaller RBs. Their RBs were similar for the different simulation choices of $n_{t,a}, \sigma_e$, and $\delta$. The RBs of the BC estimator of \cite{yu2012conditional} ($\rm{BC}_{\rm Yu}$) are generally small, although in some scenarios they can reach as high as 9\%. The BC estimator in Eq. (\ref{EQ:BC_GAM_December_10_2025}) ($\rm{BC}_{\rm GAM}$) yields large RBs when $\delta$ is small.

\begin{figure}[!h]
	\centering
	\includegraphics[width=1\linewidth]{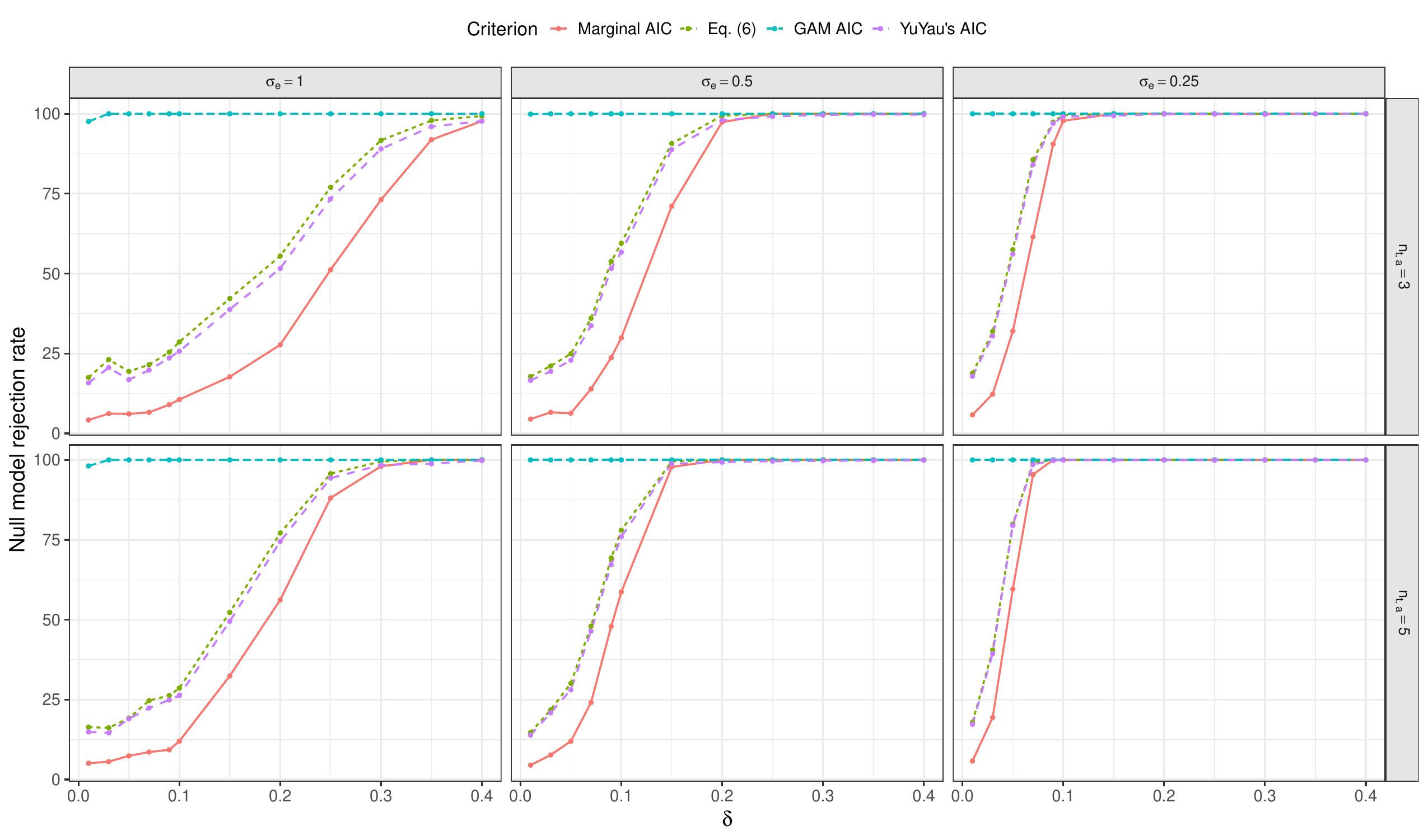}
	\caption{Rejection rates (in percentage) for the null model of no age-year interaction ($\delta=0$) under various scenarios for the Gaussian observation example. ``GAM AIC" denotes the cAIC based on Eq. (\ref{EQ:BC_GAM_December_10_2025}), and ``YuYau's AIC" denotes the cAIC proposed by \citet{yu2012conditional}.}
	\label{fig:model_selection_Gaussian_December_11_2025} 
\end{figure}

\begin{table}[ht]
	\centering
	\caption{Relative biases (RBs) in estimating the bias correction for the conditional Akaike information. The data are simulated using the linear predictor (\ref{EQ:RE_model_October_21_2024}) along with the Gaussian observational model (\ref{EQ:yi_October_21_2024}). $\rm{BC}_{\rm Yu}$ denotes the BC estimator proposed by \citet{yu2012conditional}, and $\rm{BC}_{\rm GAM}$ denotes the BC estimator given in Eq. (\ref{EQ:BC_GAM_December_10_2025}).}
	\label{tab:Gaussian_simulation_October_25_2024}
	\hspace*{-2.3cm}
	\begin{tabular}{ccccccc|ccccccc}
		\toprule
		\multicolumn{3}{c}{Design parameters} &
		\multicolumn{4}{c|}{Relative biases} &
		\multicolumn{3}{c}{Design parameters} &
		\multicolumn{4}{c}{Relative biases} \\
		\cmidrule(lr){1-3}\cmidrule(lr){4-7}
		\cmidrule(lr){8-10}\cmidrule(lr){11-14}
		$n_{t,a}$ & $\sigma_e$ & $\delta$ &
		Eq.~(\ref{EQ:CAIC_October_18_2024}) &
		Eq.~(\ref{EQ:CAIC_October_15_2024}) &
		$\rm{BC}_{\rm Yu}$ &
		$\rm{BC}_{\rm GAM}$ &
		$n_{t,a}$ & $\sigma_e$ & $\delta$ &
		Eq.~(\ref{EQ:CAIC_October_18_2024}) &
		Eq.~(\ref{EQ:CAIC_October_15_2024}) &
		$\rm{BC}_{\rm Yu}$ &
		$\rm{BC}_{\rm GAM}$ \\ 
		\midrule
		3 & 1 & 0.01 & 0.003 & 0.025 & -0.042 & -0.744 & 5 & 1 & 0.01 & -0.006 & 0.019 & -0.030 & -1.613 \\ 
		3 & 1 & 0.10 & 0.013 & 0.023 & -0.090 & -2.160 & 5 & 1 & 0.10 & 0.013 & 0.018 & -0.038 & -1.849 \\ 
		3 & 1 & 0.40 & 0.021 & 0.007 & -0.040 & -0.051 & 5 & 1 & 0.40 & 0.025 & -0.004 & -0.013 & -0.009 \\ 
		3 & 1 & 0.80 & 0.017 & -0.001 & -0.031 & -0.000 & 5 & 1 & 0.80 & 0.023 & -0.003 & -0.003 & 0.003 \\ 
		3 & 0.5 & 0.01 & 0.025 & -0.002 & -0.018 & -5.752 & 5 & 0.5 & 0.01 & 0.020 & 0.001 & -0.023 & -1.555 \\ 
		3 & 0.5 & 0.10 & 0.003 & -0.008 & 0.007 & -0.914 & 5 & 0.5 & 0.10 & 0.014 & 0.031 & -0.021 & -0.332 \\ 
		3 & 0.5 & 0.40 & 0.023 & 0.006 & 0.065 & 0.001 & 5 & 0.5 & 0.40 & 0.022 & 0.012 & 0.027 & 0.005 \\ 
		3 & 0.5 & 0.80 & 0.023 & 0.003 & -0.014 & -0.008 & 5 & 0.5 & 0.80 & 0.026 & -0.007 & -0.005 & 0.002 \\ 
		3 & 0.25 & 0.01 & 0.010 & 0.002 & -0.026 & -2.036 & 5 & 0.25 & 0.01 & 0.008 & 0.002 & -0.018 & -1.955 \\ 
		3 & 0.25 & 0.10 & 0.023 & 0.006 & -0.008 & -0.036 & 5 & 0.25 & 0.10 & 0.020 & 0.010 & -0.010 & -0.011 \\ 
		3 & 0.25 & 0.40 & 0.022 & -0.006 & -0.053 & -0.000 & 5 & 0.25 & 0.40 & 0.026 & 0.002 & -0.003 & 0.004 \\ 
		3 & 0.25 & 0.80 & 0.019 & -0.001 & -0.009 & -0.003 & 5 & 0.25 & 0.80 & 0.027 & 0.001 & -0.000 & 0.006 \\ 
		\bottomrule
	\end{tabular}
\end{table}
\FloatBarrier
The model selection results are presented in Fig. \ref{fig:model_selection_Gaussian_December_11_2025}. The rejection rates for the null model of no age-year interaction ($\delta=0$) are fairly similar for the cAIC estimator in Eq. (\ref{EQ:CAIC_October_18_2024}) and that of \citet{yu2012conditional} (``YuYau's AIC"). These results also exhibit the typical behavior of cAIC in model selection: for $\delta$ near zero, cAIC yields rejection rates of roughly 16\% for comparison of models differing in 1 parameter \citep{wood2016smoothing}, in contrast to the near-zero rejection rates of marginal AIC; as $\delta$ increases, the rejection rates for both AIC and cAIC rise, with cAIC consistently producing higher rejection rates \citep{yu2012conditional,saefken2014unifying,wood2016smoothing}. In contrast, the rejection rate for the cAIC based on Eq. (\ref{EQ:BC_GAM_December_10_2025}) (``GAM AIC") deviates markedly from this pattern, remaining close to 1 across all $\delta$ values. Note that in the lower-left panel, when $\delta>0.3$, the rejection rate for ``YuYau's AIC" is slightly lower than that of the marginal AIC, whereas the estimator in Eq.~(\ref{EQ:CAIC_October_18_2024}) does not exhibit this anomaly.

\FloatBarrier

\subsubsection{Gamma observations}
\label{sec:gamma_October_21_2024}

We assume that, conditional on all the REs, the observations follow independent gamma distributions.
For the gamma shape parameter, we examine a value of 3 for the skewed case and a value of 5 for the near-Gaussian case. The mean $\mu_{t,a}$ of the observations at year $t$ and age $a$ is modeled using a log-link function,
\begin{linenomath*} 
	\begin{align}\label{EQ:log_link_NB_October_28_2024}
		\begin{split}
			\log( \mu_{t,a} ) &=\eta_{t,a}.
		\end{split}
	\end{align}
\end{linenomath*}  
The total number of years is $T=50$. When applying Eq. (\ref{EQ:CAIC_October_18_2024}), we set $n_{\mathrm{out}}=2500$ and $n_{\mathrm{inner}}=20000$. For Eq. (\ref{EQ:CAIC_October_15_2024}), due to the higher computational cost of evaluating the derivatives in Eq. (\ref{EQ:derivatives_hattheta_Psi_wrt_y_October_10_2024}), we used $n_{\mathrm{out}}=1000$ and $n_{\mathrm{inner}}=10000$.

The BC evaluation results are shown in Table \ref{tab:Gamma_simulation_October_28_2024}. Method (\ref{EQ:CAIC_October_15_2024}) performed well with small RBs in the near-Gaussian case (shape = 5). However, it exhibited large RBs in the skewed case (shape = 3) when $\delta = 0.01$ or $0.1$, suggesting that higher-order derivatives with respect to $y$ may be necessary. In contrast, method (\ref{EQ:CAIC_October_18_2024}) performed reliably across all levels of skewness in the observational distribution, showing small RBs across all simulation scenarios. In the Supplementary Material, we provide simulation results for an alternative scenario involving gamma observations and a more complex link function, demonstrating that (\ref{EQ:CAIC_October_15_2024}) can fail for skewed distributions with intricate link functions, whereas (\ref{EQ:CAIC_October_18_2024}) exhibits significantly greater robustness. $\rm{BC}_{\rm Yu}$ exhibits slightly inferior performance compared to Eq.~(\ref{EQ:CAIC_October_18_2024}), yet it substantially outperforms both Eq.~(\ref{EQ:CAIC_October_15_2024}) and $\rm{BC}_{\rm GAM}$, with $\rm{BC}_{\rm GAM}$ continuing to fail for small values of $\delta$.

\begin{table}[ht]
	\centering
	\caption{Relative biases (RBs) in estimating bias correction for the conditional Akaike information, based on data simulated with the linear predictor (\ref{EQ:RE_model_October_21_2024}) and gamma observational distribution with log-link function (\ref{EQ:log_link_NB_October_28_2024}). $\rm{BC}_{\rm Yu}$ denotes the BC estimator proposed by \citet{yu2012conditional}, and $\rm{BC}_{\rm GAM}$ denotes the BC estimator given in Eq. (\ref{EQ:BC_GAM_December_10_2025}).}
	\label{tab:Gamma_simulation_October_28_2024}
	\hspace*{-2.3cm}
	\begin{tabular}{ccccccc|ccccccc}
		\toprule
		\multicolumn{3}{c}{Design parameters} &
		\multicolumn{4}{c|}{Relative biases} &
		\multicolumn{3}{c}{Design parameters} &
		\multicolumn{4}{c}{Relative biases} \\
		\cmidrule(lr){1-3}\cmidrule(lr){4-7}
		\cmidrule(lr){8-10}\cmidrule(lr){11-14}
		$n_{t,a}$ & Shape & $\delta$ & Eq.~(\ref{EQ:CAIC_October_18_2024}) & Eq.~(\ref{EQ:CAIC_October_15_2024}) & $\rm{BC}_{\rm Yu}$ &
		$\rm{BC}_{\rm GAM}$ & $n_{t,a}$ & Shape & $\delta$ & Eq.~(\ref{EQ:CAIC_October_18_2024}) & Eq.~(\ref{EQ:CAIC_October_15_2024}) & $\rm{BC}_{\rm Yu}$ &
		$\rm{BC}_{\rm GAM}$ \\ 
		\midrule
		3 & 3 & 0.01 & 0.006 & 0.149 & -0.031 & -2.079 & 5 & 3 & 0.01 & -0.002 & 0.183 & -0.035 & -1.431 \\ 
		3 & 3 & 0.10 & -0.016 & 0.160 & -0.039 & -1.283 & 5 & 3 & 0.10 & -0.008 & 0.137 & -0.002 & -0.711 \\ 
		3 & 3 & 0.40 & -0.042 & 0.018 & -0.017 & -0.071 & 5 & 3 & 0.40 & -0.028 & 0.018 & -0.055 & -0.046 \\ 
		3 & 3 & 0.80 & -0.077 & -0.059 & -0.099 & -0.097 & 5 & 3 & 0.80 & -0.034 & -0.013 & -0.069 & -0.063 \\ 
		3 & 5 & 0.01 & 0.021 & 0.049 & -0.041 & -2.226 & 5 & 5 & 0.01 & 0.009 & 0.067 & -0.011 & -1.600 \\ 
		3 & 5 & 0.10 & 0.007 & 0.046 & -0.033 & -1.085 & 5 & 5 & 0.10 & -0.007 & 0.060 & -0.012 & -0.208 \\ 
		3 & 5 & 0.40 & -0.032 & -0.022 & -0.058 & -0.049 & 5 & 5 & 0.40 & -0.007 & -0.007 & -0.035 & -0.027 \\ 
		3 & 5 & 0.80 & -0.042 & -0.050 & -0.071 & -0.065 & 5 & 5 & 0.80 & -0.017 & -0.031 & -0.038 & -0.035 \\  
		\bottomrule
	\end{tabular}
\end{table}

In the current design of Eq.~(\ref{EQ:RE_model_October_21_2024}), the RE distribution includes only the correlation parameter $\rho$. \cite{yu2012conditional} explicitly account for uncertainty in the estimation of the RE-distribution parameters. To evaluate the performance of their approach when additional RE-distribution parameters are present, we revise Eq.~(\ref{EQ:RE_model_October_21_2024}) by including the sd parameters $\sigma$ and $\delta$ in the RE distribution. Specifically, we set
\begin{linenomath*} 
	\begin{align}\label{EQ:RE_model_more_RE_parameters_December_12_2025}
		\begin{split}
			\eta_{t,a} &= q_a + \Psi_t + \Psi_{t,a},
		\end{split}
	\end{align}
\end{linenomath*} 
where $\Psi_t$ follows an AR(1) process with marginal sd $\sigma$ and $\Psi_{t,a}$ are independent normal errors with sd $\delta$. This modification yields a total of three parameters in the RE distribution. The simulation results, summarized in Table \ref{tab:Gamma_simulation_more_RE_parameters_December_7_2025} of the Supplementary Material, show that $\rm{BC}_{\rm Yu}$ performs poorly, particularly for small $\delta$, whereas Eq.~(\ref{EQ:CAIC_October_18_2024}) continues to perform well. The performance of $\rm{BC}_{\rm GAM}$ also improves noticeably. In the following simulation studies we continue to use the linear predictor in Eq. (\ref{EQ:RE_model_October_21_2024}) rather than that of Eq. (\ref{EQ:RE_model_more_RE_parameters_December_12_2025}).

\begin{figure}[!h]
	\centering
	\includegraphics[width=1\linewidth]{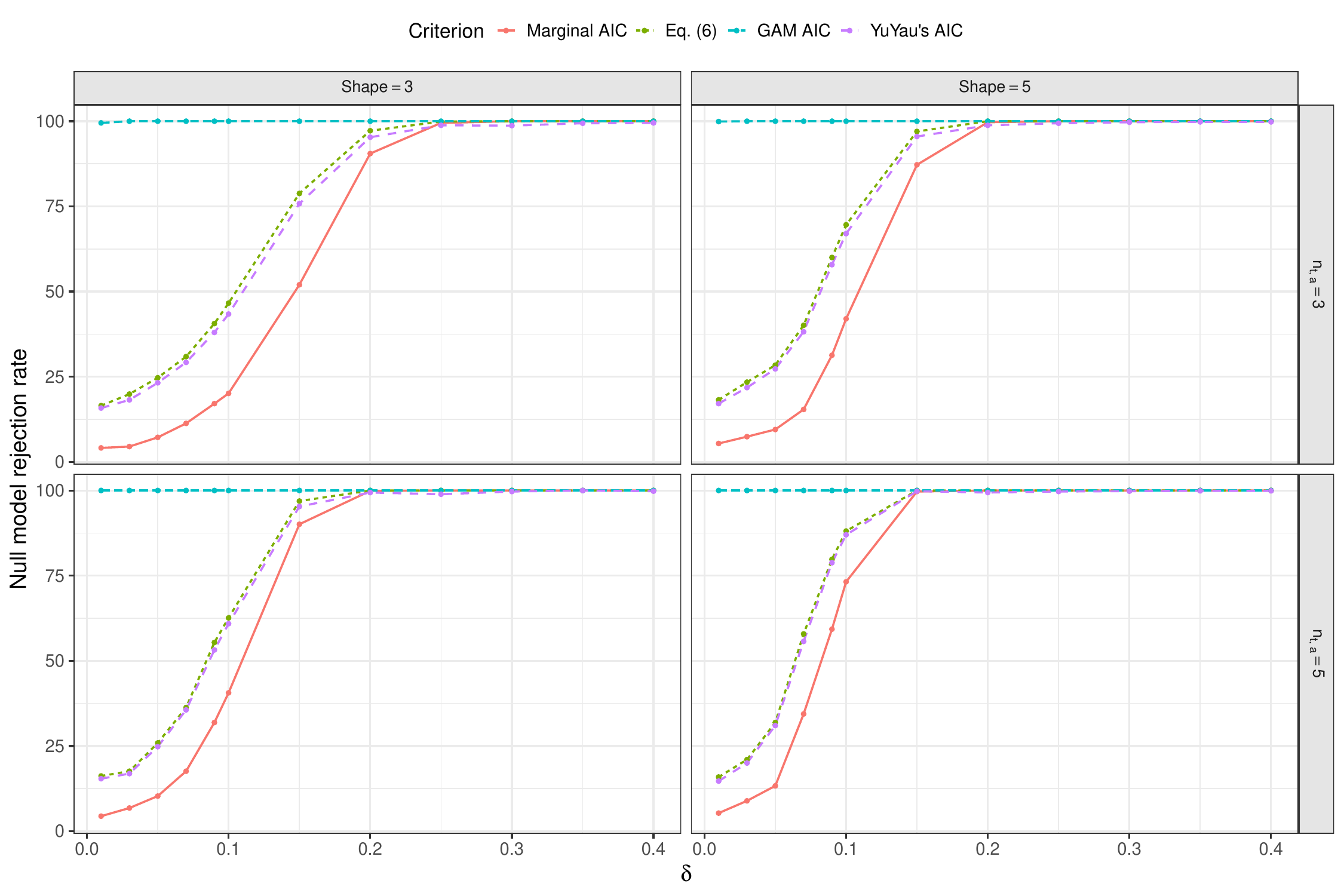}
	\caption{Rejection rates (in percentage) for the null model of no age-year interaction ($\delta = 0$) across scenarios in the gamma-log link example. ``GAM AIC" denotes the cAIC based on Eq. (\ref{EQ:BC_GAM_December_10_2025}), and ``YuYau's AIC" denotes the cAIC proposed by \citet{yu2012conditional}.}
	\label{fig:Model_selection_gamma_log_link_December_12_2025} 
\end{figure}

\FloatBarrier
The model selection results presented in Fig.~\ref{fig:Model_selection_gamma_log_link_December_12_2025} exhibit a pattern similar to that in Fig.~\ref{fig:model_selection_Gaussian_December_11_2025}. For larger values of $\delta$, ``YuYau's AIC" again yields slightly lower rejection rates than the marginal AIC, particularly in the two left panels.

\FloatBarrier
\subsubsection{Negative binomial observations}
\label{sec:negative_binomial_October_28_2024}
We assume that, given all the REs, the observations are distributed independently with negative binomial distributions. We apply the log-link function (\ref{EQ:log_link_NB_October_28_2024}) to model the mean $\mu_{t,a}$ of the observations at year $t$ and age $a$. The corresponding variance is defined by the dispersion relationship of the negative binomial distribution, $\mu_{t,a} + \alpha\, \mu_{t,a}^2$, where the dispersion parameter $\alpha$ is set to be 0.2 or 0.5. The total number of years is $T=50$. Because the data follow a discrete distribution, we apply only the proposed method (\ref{EQ:CAIC_October_18_2024}), with $n_{\mathrm{out}}=500$ and $n_{\mathrm{inner}}=1000$. The simulation results in Table \ref{tab:negative_binomial_simulation_October_28_2024} show that method (\ref{EQ:CAIC_October_18_2024}) achieved low RBs across all simulation scenarios. Overall, $\rm{BC}_{\rm Yu}$ performed well, except in a few cases with RB values of -0.123 and 0.088, while $\rm{BC}_{\rm GAM}$ still failed for small $\delta$.

\begin{table}[ht]
	\centering
	\caption{Relative biases (RBs) in estimating bias correction for the conditional Akaike Information, based on data simulated with the linear predictor in Eq. (\ref{EQ:RE_model_October_21_2024}) and negative binomial observational distribution with mean defined by Eq. (\ref{EQ:log_link_NB_October_28_2024}) and dispersion parameter $\alpha$. $\rm{BC}_{\rm Yu}$ denotes the BC estimator proposed by \citet{yu2012conditional}, and $\rm{BC}_{\rm GAM}$ denotes the BC estimator given in Eq. (\ref{EQ:BC_GAM_December_10_2025}).}
	\label{tab:negative_binomial_simulation_October_28_2024}
	\hspace*{-0.7cm}
	\begin{tabular}{cccccc|cccccc}
		\toprule
		\multicolumn{3}{c}{Design parameters} &
		\multicolumn{3}{c|}{Relative biases} &
		\multicolumn{3}{c}{Design parameters} &
		\multicolumn{3}{c}{Relative biases} \\
		\cmidrule(lr){1-3}\cmidrule(lr){4-6}
		\cmidrule(lr){7-9}\cmidrule(lr){10-12}
		$n_{t,a}$ & $\alpha$ & $\delta$ & Eq.~(\ref{EQ:CAIC_October_18_2024}) &  $\rm{BC}_{\rm Yu}$ &
		$\rm{BC}_{\rm GAM}$  & $n_{t,a}$ & $\alpha$ & $\delta$ & Eq.~(\ref{EQ:CAIC_October_18_2024}) &  $\rm{BC}_{\rm Yu}$ &
		$\rm{BC}_{\rm GAM}$  \\ 
		\midrule
		3 & 0.2 & 0.01 & -0.003 & -0.017 & -1.143 & 5 & 0.2 & 0.01 & 0.001 & -0.004 & -1.420 \\ 
		3 & 0.2 & 0.10 & 0.022 & -0.039 & -1.034 & 5 & 0.2 & 0.10 & 0.028 & -0.023 & -1.194 \\ 
		3 & 0.2 & 0.40 & -0.017 & -0.017 & -0.026 & 5 & 0.2 & 0.40 & 0.005 & 0.016 & -0.013 \\ 
		3 & 0.2 & 0.80 & 0.011 & -0.016 & -0.007 & 5 & 0.2 & 0.80 & 0.017 & -0.008 & 0.001 \\ 
		3 & 0.5 & 0.01 & -0.004 & -0.088 & -0.679 & 5 & 0.5 & 0.01 & -0.007 & -0.043 & -0.799 \\ 
		3 & 0.5 & 0.10 & -0.026 & -0.123 & -1.948 & 5 & 0.5 & 0.10 & 0.048 & -0.049 & -1.438 \\ 
		3 & 0.5 & 0.40 & -0.027 & -0.065 & -0.086 & 5 & 0.5 & 0.40 & -0.011 & -0.038 & -0.038 \\ 
		3 & 0.5 & 0.80 & -0.045 & -0.065 & -0.062 & 5 & 0.5 & 0.80 & -0.020 & -0.054 & -0.047 \\ 
		\bottomrule
	\end{tabular}
\end{table}

\begin{figure}[!h]
	\centering
	\includegraphics[width=1\linewidth]{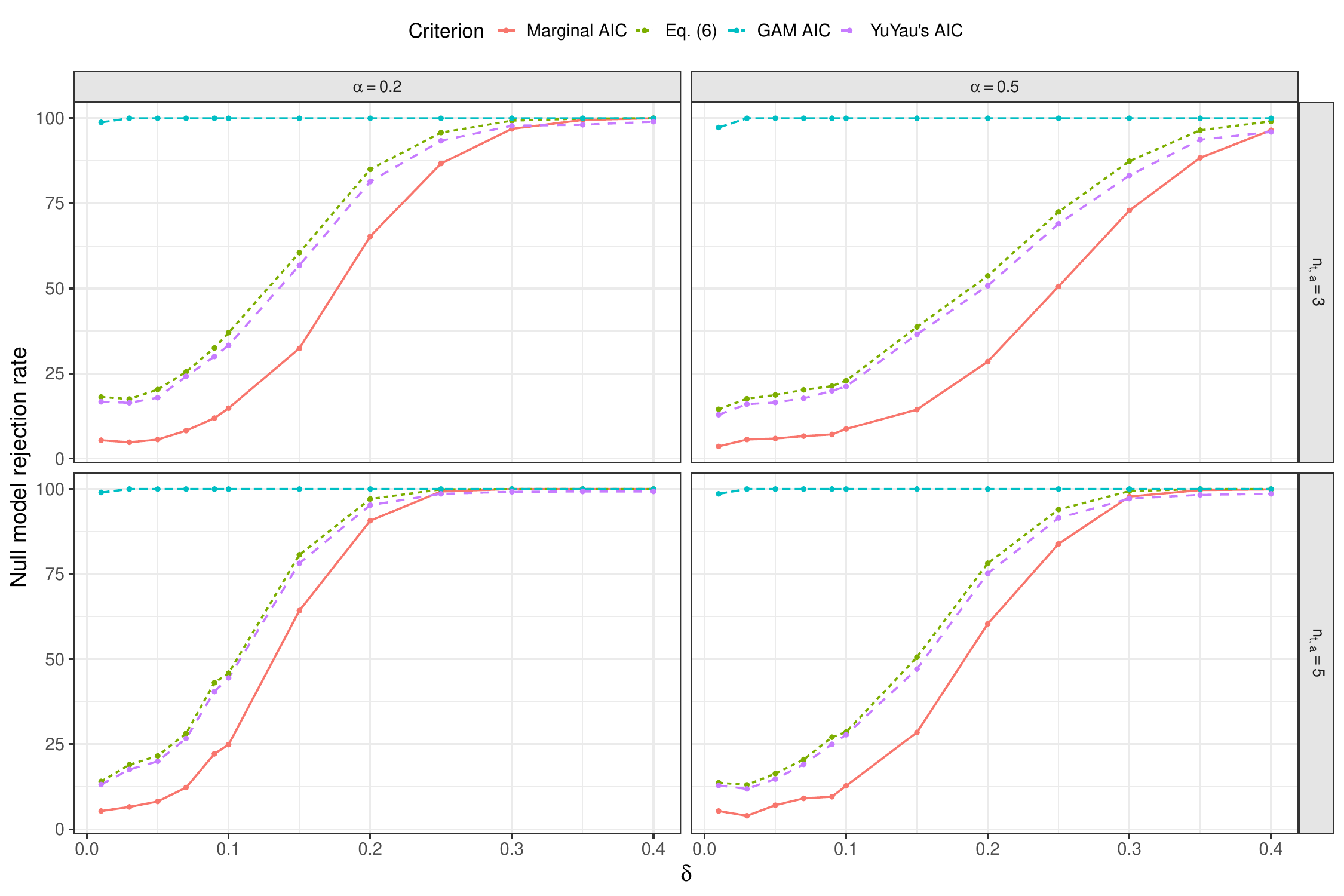}
	\caption{Rejection rates (in percentage) for the null model of no age-year interaction ($\delta = 0$) across scenarios in the negative binomial example. ``GAM AIC" denotes the cAIC based on Eq. (\ref{EQ:BC_GAM_December_10_2025}), and ``YuYau's AIC" denotes the cAIC proposed by \citet{yu2012conditional}.}
	\label{fig:model_selection_NB_December_13_2025} 
\end{figure}

\FloatBarrier
The model selection results shown in Fig.~\ref{fig:model_selection_NB_December_13_2025} convey information consistent with that in Figs.~\ref{fig:model_selection_Gaussian_December_11_2025}--\ref{fig:Model_selection_gamma_log_link_December_12_2025}. In particular, Eq.~(\ref{EQ:CAIC_October_18_2024}) and ``YuYau's AIC" exhibit similar null-model rejection rates; however, for larger values of $\delta$, the rejection rate of ``YuYau's AIC" can be lower than that of the marginal AIC.

\FloatBarrier

\subsubsection{Tweedie observations}

The Tweedie distribution \citep{jorgensen1997theory} encompasses a family of important distributions indexed by the power parameter $p$. To evaluate the performance of method (\ref{EQ:CAIC_October_18_2024}) for zero-inflated distributions, we focus on the case $1<p<2$, where the Tweedie distribution is a compound Poisson-gamma distribution that is continuous over the positive real axis and includes a point mass at zero. Specifically, we set $p=1.5$ and use dispersion parameter values $\phi=0.3$ or 0.5, applying the log-link function (\ref{EQ:log_link_NB_October_28_2024}) to model the mean. The variance of the Tweedie distribution follows the power law $\mathrm{Var}(Y_{t,a})=\phi\, \mu_{t,a}^p$. Because the observation distribution lacks differentiability across its entire support, we apply only the proposed method (\ref{EQ:CAIC_October_18_2024}). Additionally, due to the relatively slow generation of Tweedie random variables, we set $T=30$, $n_{\mathrm{out}} = 500$ and $n_{\mathrm{inner}} = 1000$. The simulation results in Table \ref{tab:Tweedie_simulation_October_28_2024} indicate that method (\ref{EQ:CAIC_October_18_2024}) achieved low RBs for Tweedie observations across all scenarios. $\rm{BC}_{\rm Yu}$ also performed well, whereas $\rm{BC}_{\rm GAM}$ failed when $\delta$ was small.

\begin{table}[ht]
	\centering
	\caption{Relative biases (RBs) in estimating bias correction for the conditional Akaike Information, based on data simulated with the linear predictor in Eq. (\ref{EQ:RE_model_October_21_2024}) and Tweedie observational distribution with mean defined by Eq. (\ref{EQ:log_link_NB_October_28_2024}), power parameter $p=1.5$ and dispersion parameter $\phi$. $\rm{BC}_{\rm Yu}$ denotes the BC estimator proposed by \citet{yu2012conditional}, and $\rm{BC}_{\rm GAM}$ denotes the BC estimator given in Eq. (\ref{EQ:BC_GAM_December_10_2025}).}
	\label{tab:Tweedie_simulation_October_28_2024}
	\hspace*{-0.7cm}
	\begin{tabular}{cccccc|cccccc}
		\toprule
		\multicolumn{3}{c}{Design parameters} &
		\multicolumn{3}{c|}{Relative biases} &
		\multicolumn{3}{c}{Design parameters} &
		\multicolumn{3}{c}{Relative biases} \\
		\cmidrule(lr){1-3}\cmidrule(lr){4-6}
		\cmidrule(lr){7-9}\cmidrule(lr){10-12}
		$n_{t,a}$ & $\phi$ & $\delta$ & Eq.~(\ref{EQ:CAIC_October_18_2024}) &  $\rm{BC}_{\rm Yu}$ &
		$\rm{BC}_{\rm GAM}$  & $n_{t,a}$ & $\phi$ & $\delta$ & Eq.~(\ref{EQ:CAIC_October_18_2024}) &  $\rm{BC}_{\rm Yu}$ &
		$\rm{BC}_{\rm GAM}$  \\ 
		\midrule
		3 & 0.3 & 0.01 & 0.014 & 0.070 & -1.213 & 5 & 0.3 & 0.01 & 0.031 & -0.076 & -0.924 \\ 
		3 & 0.3 & 0.10 & -0.018 & -0.008 & -0.531 & 5 & 0.3 & 0.10 & 0.021 & -0.045 & -0.162 \\ 
		3 & 0.3 & 0.40 & 0.005 & -0.017 & -0.008 & 5 & 0.3 & 0.40 & 0.023 & -0.014 & -0.005 \\ 
		3 & 0.3 & 0.80 & 0.019 & -0.042 & -0.035 & 5 & 0.3 & 0.80 & 0.030 & -0.021 & -0.012 \\ 
		3 & 0.5 & 0.01 & 0.016 & -0.050 & -1.736 & 5 & 0.5 & 0.01 & 0.014 & -0.002 & -1.221 \\ 
		3 & 0.5 & 0.10 & 0.012 & -0.023 & -0.880 & 5 & 0.5 & 0.10 & -0.004 & -0.032 & -0.648 \\ 
		3 & 0.5 & 0.40 & -0.004 & -0.034 & -0.028 & 5 & 0.5 & 0.40 & 0.017 & -0.031 & -0.020 \\ 
		3 & 0.5 & 0.80 & -0.001 & -0.044 & -0.037 & 5 & 0.5 & 0.80 & 0.023 & -0.010 & -0.002 \\ 
		\bottomrule
	\end{tabular}
\end{table}

\begin{figure}[!h]
	\centering
	\includegraphics[width=1\linewidth]{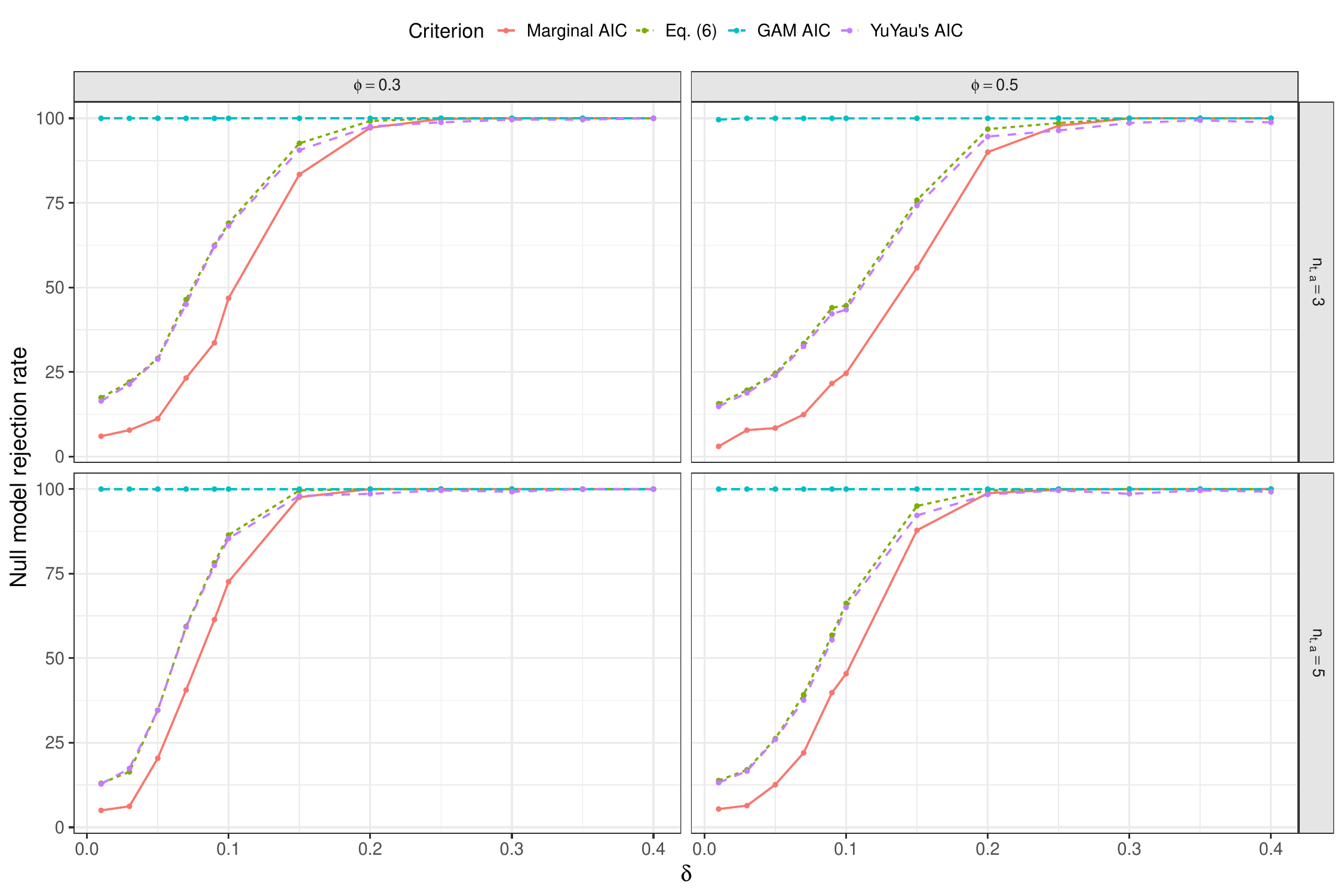}
	\caption{Rejection rates (in percentage) for the null model of no age-year interaction ($\delta = 0$) across scenarios in the Tweedie example. ``GAM AIC" denotes the cAIC based on Eq. (\ref{EQ:BC_GAM_December_10_2025}), and ``YuYau's AIC" denotes the cAIC proposed by \citet{yu2012conditional}.}
	\label{fig:model_selection_Tweedie_December_13_2025} 
\end{figure}

\FloatBarrier
The model selection results in Fig.~\ref{fig:model_selection_Tweedie_December_13_2025} exhibit the same pattern as those in Figs.~\ref{fig:model_selection_Gaussian_December_11_2025}--\ref{fig:model_selection_NB_December_13_2025}. When $\delta$ is close to 0, the rejection rates based on the marginal AIC are close to zero, whereas those based on Eq.~(\ref{EQ:CAIC_October_18_2024}) and ``YuYao's AIC" are similar and range from approximately 13\% to 17\%. For larger values of $\delta$, the rejection rate of ``YuYao's AIC" can fall below that of the marginal AIC. The rejection rate of ``GAM AIC" is always close to 1.

\FloatBarrier

\subsection{Three-dimensional spatiotemporal model}
We consider a spatiotemporal model to examine the performance of the proposed cAIC evaluation method under more complicated RE models. The linear predictor becomes
\begin{linenomath*} 
	\begin{align}\label{EQ:3D_RE_model_December_14_2025}
		\begin{split}
			\eta_{t,x,y} &= q + \sigma\, \Psi_{x,y} + \delta\, \Psi_{t,x,y},
		\end{split}
	\end{align}
\end{linenomath*} 
where $q=1$, $\sigma=1$ and $\delta$ are parameters, $\Psi_{x,y}$ is an RE following a 2-dimensional Gaussian Markov random field (GMRF) with marginal variance equal to 1 and autocorrelations along $x$ and $y$ directions equal to 0.6, and $\Psi_{t,x,y}$ is a 3-dimensional (3D) GMRF with temporal autocorrelation (along $t$) being 0.8 and spatial autocorrelation (along $x$ and $y$) being 0.7.

Conditional on all REs, the observations are assumed to be independently distributed according to negative binomial distributions. A log-link function, $\log(\mu_{t,x,y}) = \eta_{t,x,y}$, is used to relate the mean $\mu_{t,x,y}$ of the observations at year $t$ and spatial grid unit indexed by $(x,y)$ to the linear predictor. The corresponding variance follows the negative binomial dispersion relationship, $\mu_{t,x,y} + \alpha\,\mu_{t,x,y}^2$, where the dispersion parameter $\alpha$ is set to 0.2 or 0.5. To maintain computational feasibility, a relatively small spatiotemporal grid is considered: the total number of years is 8, and the numbers of grid units in both the $x$ and $y$ directions are 8. The number of observations per spatiotemporal grid unit is set to $n_{t,x,y}=2$ or 4. Because $\rm{BC}_{\rm GAM}$ performed poorly in the preceding simulation studies, we restrict the comparison to Eq.~(\ref{EQ:CAIC_October_18_2024}) and $\rm{BC}_{\rm Yu}$, using $n_{\mathrm{out}}=500$ and $n_{\mathrm{inner}}=1000$.

The simulation results in Table~\ref{tab:3D_negative_binomial_simulation_December_16_2025} indicate that method~(\ref{EQ:CAIC_October_18_2024}) yields consistently low RBs across all simulation scenarios. In contrast, $\mathrm{BC}_{\mathrm{Yu}}$ exhibits inferior overall performance, with notably larger RBs in several cases, reaching values of $-0.125$ and $-0.186$.

\begin{table}[ht]
	\centering
	\caption{Relative biases (RBs) of bias-correction estimators for the conditional Akaike Information, based on data simulated from the linear predictor in Eq.~(\ref{EQ:3D_RE_model_December_14_2025}) with a negative binomial observation model using a log link and dispersion parameter $\alpha$. $\rm{BC}_{\rm Yu}$ denotes the bias-correction estimator proposed by \citet{yu2012conditional}.}
	\label{tab:3D_negative_binomial_simulation_December_16_2025}
	\begin{tabular}{ccccc|ccccc}
		\toprule
		\multicolumn{3}{c}{Design parameters} &
		\multicolumn{2}{c|}{Relative biases} &
		\multicolumn{3}{c}{Design parameters} &
		\multicolumn{2}{c}{Relative biases} \\
		\cmidrule(lr){1-3}\cmidrule(lr){4-5}
		\cmidrule(lr){6-8}\cmidrule(lr){9-10}
		$n_{t,x,y}$ & $\alpha$ & $\delta$ & Eq.~(\ref{EQ:CAIC_October_18_2024}) &  $\rm{BC}_{\rm Yu}$ & $n_{t,x,y}$ & $\alpha$ & $\delta$ & Eq.~(\ref{EQ:CAIC_October_18_2024}) &  $\rm{BC}_{\rm Yu}$ \\
		\midrule
		2 & 0.2 & 0.01 & -0.004 & -0.005 
		& 4 & 0.2 & 0.01 &  0.003 & -0.048 \\
		2 & 0.2 & 0.10 & -0.019 & -0.058 
		& 4 & 0.2 & 0.10 &  0.007 & -0.035 \\
		2 & 0.2 & 0.40 & -0.002 & -0.028 
		& 4 & 0.2 & 0.40 & -0.033 & -0.058 \\
		2 & 0.2 & 0.80 & -0.015 & -0.028 
		& 4 & 0.2 & 0.80 &  0.004 & -0.010 \\
		2 & 0.5 & 0.01 & -0.022 & -0.069 
		& 4 & 0.5 & 0.01 & -0.016 & -0.186 \\
		2 & 0.5 & 0.10 & -0.022 & -0.070 
		& 4 & 0.5 & 0.10 & -0.009 & -0.053 \\
		2 & 0.5 & 0.40 & -0.020 & -0.125 
		& 4 & 0.5 & 0.40 & -0.015 & -0.032 \\
		2 & 0.5 & 0.80 & -0.023 & -0.044 
		& 4 & 0.5 & 0.80 & -0.033 & -0.047 \\ 
		\bottomrule
	\end{tabular}
\end{table}

\begin{figure}[!h]
	\centering
	\includegraphics[width=1\linewidth]{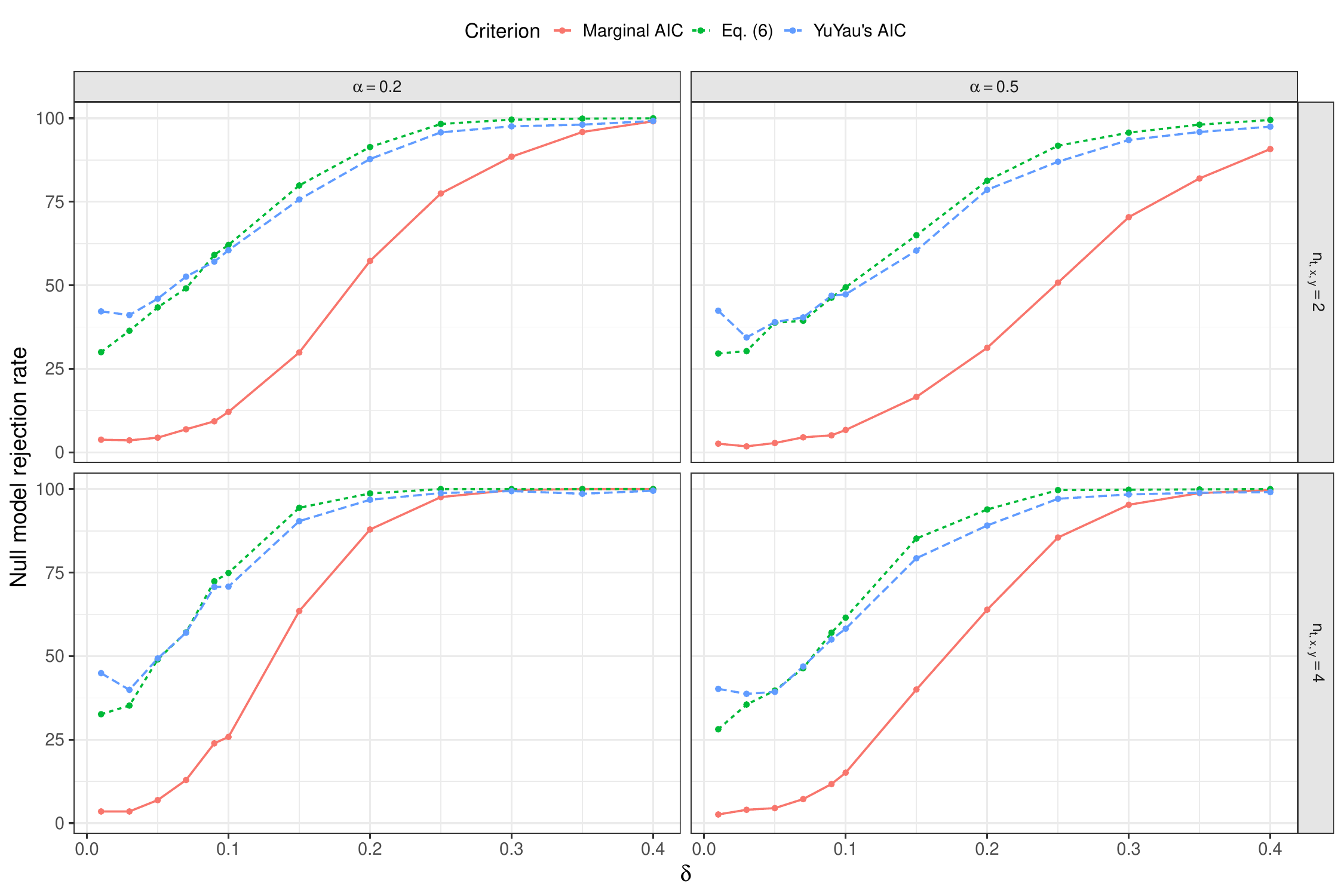}
	\caption{Rejection rates (in percentage) of the null model with no 3D GMRF effect ($\delta = 0$) in model (\ref{EQ:3D_RE_model_December_14_2025}) across simulation scenarios for the 3D spatiotemporal example. ``YuYau's AIC" denotes the cAIC proposed by \citet{yu2012conditional}.}
	\label{fig:model_selection_spatiotemporal_3D_December_22_2025} 
\end{figure}

\FloatBarrier
In Fig.~\ref{fig:model_selection_spatiotemporal_3D_December_22_2025}, $\mathrm{BC}_{\mathrm{Yu}}$ generally follows Eq.~(\ref{EQ:CAIC_October_18_2024}), but exhibits an anomaly, with higher null-model rejection rates for smaller $\delta$ than for larger $\delta$. This behavior is not observed for the cAIC estimator in Eq.~(\ref{EQ:CAIC_October_18_2024}). For small $\delta$, the marginal AIC strongly favors the null model, whereas for larger $\delta$ the rejection rate of ``Yu-Yau's AIC" can fall below that of the marginal AIC (e.g., lower-left panel).
 
\FloatBarrier
\subsection{Comparison with the R package \texttt{cAIC4}}\label{compare_cAIC4_August_21_2026}
The \texttt{cAIC4} package currently supports cAIC evaluation only for Gaussian, Poisson, and binomial responses. For the Gaussian case, ``exact" expressions for cAIC are available in the literature and are expected to yield performance comparable to that of Eq.~(\ref{EQ:CAIC_October_15_2024}). Consequently, we do not include comparisons with \texttt{cAIC4} for Gaussian responses. Instead, we restrict the comparison to the Poisson and binomial cases, for which \texttt{cAIC4} is applicable.

The \texttt{cAIC4} package computes cAIC only for generalized mixed-effects models fitted with \texttt{(g)lmer()} from \texttt{lme4}, \texttt{lme()} from \texttt{nlme}, and \texttt{gamm()} from \texttt{mgcv}. These implementations do not accommodate AR(1) structures in the REs. Moreover, the computational cost of \texttt{cAIC4} is substantial, limiting its practical use to models significantly smaller than those previously considered. We therefore adopt a simplified linear predictor, compared to Eq.~(\ref{EQ:RE_model_October_21_2024}),
\begin{linenomath*} 
	\begin{align}\label{EQ:simpler_RE_model_April_11_2026}
		\begin{split}
			\eta_{t,a} &= q_a + \delta\, \Psi_{t,a},
		\end{split}
	\end{align}
\end{linenomath*} 	
where $q_a$ and $\delta$ are model parameters. There are a total of 20 years with $t=1,2,\ldots,20$. Each year includes 3 age groups indexed by subscript $a$, with $q_a=(-2, 0, 2)$. The REs, $\Psi_{t,a}$, are independent standard normal variables.
\subsubsection{Poisson observations}
For Poisson observations, the log-link specified in Eq.~(\ref{EQ:log_link_NB_October_28_2024}) is employed. To implement \texttt{cAIC4}, the model is fitted to the simulated data using the \texttt{glmer()} function from the \texttt{lme4} package, after which the cAIC is computed via the \texttt{cAIC()} function in \texttt{cAIC4}. The simulation results are presented in Table~\ref{tab:Poisson_cAIC4_April_10_2026}. The \texttt{cAIC4} procedure fails to produce results for the case $\delta=0.01$; accordingly, this scenario is omitted.	
\begin{table}[ht]
	\centering
	\caption{Relative biases (RBs) in estimating bias correction for the conditional Akaike information, based on data simulated with the linear predictor (\ref{EQ:simpler_RE_model_April_11_2026}) and Poisson observational distribution with log-link function (\ref{EQ:log_link_NB_October_28_2024}). $\rm{BC}_{\rm cAIC4}$ denotes the BC estimated using the \texttt{cAIC()} function in the R package \texttt{cAIC4}, $\rm{BC}_{\rm Yu}$ denotes the BC estimator proposed by \citet{yu2012conditional}, and $\rm{BC}_{\rm GAM}$ denotes the BC estimator given in Eq.~(\ref{EQ:BC_GAM_December_10_2025}).}
	\label{tab:Poisson_cAIC4_April_10_2026}	\hspace*{-0.7cm}	
	\begin{tabular}{cccccc|cccccc}
		\toprule	
		\multicolumn{2}{c}{Design} & \multicolumn{4}{c|}{Relative biases} &
		\multicolumn{2}{c}{Design} & \multicolumn{4}{c}{Relative biases} \\
		\cmidrule(lr){1-2} \cmidrule(lr){3-6} \cmidrule(lr){7-8} \cmidrule(lr){9-12}
		$n_{t,a}$ & $\delta$ & Eq.~(\ref{EQ:CAIC_October_18_2024}) & $\rm{BC}_{\rm cAIC4}$ & $\rm{BC}_{\rm Yu}$ & $\rm{BC}_{\rm GAM}$ & $n_{t,a}$ & $\delta$ & Eq.~(\ref{EQ:CAIC_October_18_2024}) & $\rm{BC}_{\rm cAIC4}$ & $\rm{BC}_{\rm Yu}$ & $\rm{BC}_{\rm GAM}$\\
		\midrule
		3 & 0.1 & 0.014 & 0.172 & -0.105 & 0.157 & 5 & 0.1 & 0.043 & 0.151 & -0.080 & 0.155\\
		3 & 0.2 & -0.006 & 0.060 & -0.918 & 0.042 & 5 & 0.2 & -0.013 & 0.007 & -8.488 & 0.002\\
		3 & 0.4 & 0.009 & -0.003 & -0.141 & -0.011 & 5 & 0.4 & 0.057 & 0.030 & 0.048 & 0.026\\
		3 & 0.6 & 0.027 & -0.004 & 0.012 & -0.007 & 5 & 0.6 & 0.033 & -0.006 & 0.002 & -0.006\\
		3 & 0.8 & 0.045 & 0.005 & 0.013 & 0.005 & 5 & 0.8 & 0.063 & 0.015 & 0.024 & 0.020\\					
		\bottomrule
	\end{tabular}
\end{table}
\FloatBarrier
\subsubsection{Binomial observations}	
For binomial observations with the number of trials (size) fixed at 5, a logit link is used to model the success probabilities. The same procedure as in the Poisson case is applied to implement the \texttt{cAIC4} package for cAIC evaluation. Computation with \texttt{cAIC4} is substantially more time-consuming for binomial responses than for Poisson observations; consequently, only the case with sample size $n_{t,a}=3$ is considered. The simulation results are reported in Table~\ref{tab:binomial_cAIC4_April_12_2026}.
\begin{table}[ht]
	\centering
	\caption{Relative biases (RBs) in estimating bias correction for the conditional Akaike information, based on data simulated with the linear predictor (\ref{EQ:simpler_RE_model_April_11_2026}) and binomial observational distribution with logit-link function. $\rm{BC}_{\rm cAIC4}$ denotes the BC estimated using the \texttt{cAIC()} function in the R package \texttt{cAIC4}, $\rm{BC}_{\rm Yu}$ denotes the BC estimator proposed by \citet{yu2012conditional}, and $\rm{BC}_{\rm GAM}$ denotes the BC estimator given in Eq.~(\ref{EQ:BC_GAM_December_10_2025}).}
	\label{tab:binomial_cAIC4_April_12_2026}	%\hspace*{-0.7cm}	
	\begin{tabular}{cccccc}
		\toprule
		\multicolumn{2}{c}{Design} & \multicolumn{4}{c}{Relative biases} \\
		\cmidrule(lr){1-2} \cmidrule(lr){3-6}
		$n_{t,a}$ & $\delta$ & Eq.~(\ref{EQ:CAIC_October_18_2024}) & $\rm{BC}_{\rm cAIC4}$ & $\rm{BC}_{\rm Yu}$ & $\rm{BC}_{\rm GAM}$\\
		\midrule
		3 & 0.1 & -0.042 & -0.998 & -0.092 & 0.074\\
		3 & 0.2 & -0.043 & -1.017 & -0.105 & 0.066\\
		3 & 0.4 & -0.029 & -0.982 & -0.138 & 0.025\\
		3 & 0.6 & 0.019 & -0.999 & 0.004 & 0.025\\
		3 & 0.8 & 0.057 & -1.002 & 0.447 & 0.040\\
		\bottomrule
	\end{tabular}
\end{table}

\FloatBarrier
\subsubsection{Summary of the comparison results}
For Poisson responses, \texttt{cAIC4} generally yields accurate estimates of cAIC, except when the REs are weak (e.g., $\delta=0.1$ in Table \ref{tab:Poisson_cAIC4_April_10_2026}). For binomial observations (Table \ref{tab:binomial_cAIC4_April_12_2026}), however, \texttt{cAIC4} fails to produce valid cAIC estimates. In contrast, Eq.~(\ref{EQ:CAIC_October_18_2024}) performs well across all scenarios considered. Furthermore, \texttt{cAIC4} does not accommodate flexible correlation structures, such as AR(1), for the REs, and, due to its computational cost, is typically limited to applications involving relatively small model sizes. By contrast, Eq.~(\ref{EQ:CAIC_October_18_2024}) is applicable to a broad range of model structures and scales implementable with TMB.

\section{Fish stock assessment model selection}\label{sec:fish_stock_model_selection}
State-space models are popular when modeling time-varying ecological phenomena \citep{pedersen2011estimation}, including fish population dynamics \citep[e.g.,][]{schnute1994general,gudmundsson1994time, aanes2007estimation, maunder2011state,nielsen2014estimation,cadigan2015state,pedersen2017stochastic,perreault2020state}.
State-space fish stock assessment models combine stochastic assumptions about both observed quantities and unobserved states such as population abundance and mortality.
These models commonly integrate several sources of data relevant to fish population dynamics, including fisheries catches, abundance and other biological information (i.e., length, age, weight, maturity) from surveys, and tagging information.
Different types of data may have continuous or discrete distributions.

The Woods Hole Assessment Model \citep[WHAM,][]{stock2021woods,stock2021implementing} is an example of a SSAM that is used for fisheries management, primarily in the United States.
WHAM has options to use random effects (REs) for various model components (see \url{https://timjmiller.github.io/wham/}), including reproduction and mortality rates.
The model documentation includes many examples (i.e., vignettes) that describe various options and configurations for the model.
We ran example 6 that involved a variety of options for modelling the yearly transitions in cohort survival as REs, including independent or correlated effects.
This involved a total of 13 models, which are described in Section 7 of \url{https://timjmiller.github.io/wham/articles/ex06_NAA.html}, where AIC was used for model selection.
The details of these models are beyond the scope of this paper; however, models m2-m7 were the same as m8-m13 except that models m8-m13 included an environmental covariate effect on reproduction (i.e., recruitment) whereas m2-m7 did not have this effect.

We also computed cAIC using Eq.~(\ref{EQ:CAIC_October_18_2024}) (see Table \ref{tab:WHAM}) to investigate if there are differences in model selection compared to AIC.
Model m13 had the lowest value of AIC but model m11 also had a low value.
AIC strongly supported these two models compared to the others.
However, cAIC indicated model m5 was best.
This model included between-age correlation in survival REs, the same as m11, but m13 included more complicated between-age and -year correlations.
Hence, compared to AIC, cAIC indicated a simpler model configuration was best among those investigated, with no recruitment covariate effect and only between-age correlation in survival.

\begin{table}[ht]
	
	\centering
	
	\caption{WHAM fish stock assessment model selection results for example 6 in its model documentation. $p$ is the number of model parameters. $\Delta \mathrm{AIC}$ and $\Delta \mathrm{cAIC}$ are the differences in AIC and cAIC from the model with the lowest value.}
	
	\label{tab:WHAM}
	
	%\scalebox{0.75}{
		
		\begin{tabular}{llrr|llrr}
			
			\toprule
			
			Model & $p$ & $\Delta \mathrm{AIC}$ & $\Delta \mathrm{cAIC}$ & Model & $p$& $\Delta \mathrm{AIC}$ & $\Delta \mathrm{cAIC}$ \\ \midrule
			
			m2 & 72 & 512.2              & 686.8 & m8 & 73          & 492.1                & 690.2 \\
			
			m3 & 73 & 479  & 690.4 & m9 & 74          & 472.8                & 695.5 \\
			
			m4 & 73 & 42.3 & 4.9 & m10 & 74 & 25.4              & 10.5 \\
			
			m5 & 74 & 17.2 & 0 & m11 & 75 & 1.5    & 6.9 \\
			
			m6 & 74 & 27.8 & 12.6  & m12 &               75 & 18.7            & 17.7 \\
			
			m7 & 75 & 10.7 & 3.8 & m13 & 76 & 0    & 10.6 \\ \bottomrule
			
		\end{tabular}%}
	
\end{table}

\FloatBarrier
\section{Conclusion}
\label{sec:conc}

We proposed two estimators for the conditional Akaike information criterion (cAIC) in Eqs. (\ref{EQ:CAIC_October_15_2024}) and (\ref{EQ:CAIC_October_18_2024}). For Eq. (\ref{EQ:CAIC_October_15_2024}), we derived analytical expressions for the necessary derivatives in (\ref{EQ:derivatives_hattheta_Psi_wrt_y_October_10_2024}), which can be evaluated using advanced modeling packages such as TMB. To address potential bias inherent in the conventional AIC derivation of \cite{Akaike1973AIC} when applied to the cAIC framework, we adopt a new derivation method leading to Eq.~(\ref{EQ:CAIC_October_18_2024}). We used the relative bias (RB) in estimating the bias correction for the conditional Akaike information (cAI) to evaluate the estimator's performance. The evaluation was conducted under conditions of small, upper-bounded cluster sizes (3 or 5), a common scenario in spatiotemporal modeling across various disciplines. 

Simulation results indicate that (\ref{EQ:CAIC_October_15_2024}) performs well for Gaussian observations with an identity link function but fails with skewed distributions (e.g., gamma distributions with small shape parameters). Additionally, (\ref{EQ:CAIC_October_15_2024}) is unsuitable for observational distributions lacking well-defined derivatives across their supports, such as discrete or zero-inflated distributions. In contrast, (\ref{EQ:CAIC_October_18_2024}) consistently performs well across all observational distributions and link functions examined, and outperforms existing cAIC formulations in the literature \citep{yu2012conditional,wood2016smoothing}. Eq.~(\ref{EQ:CAIC_October_18_2024}) also exhibits the expected behavior of cAI in model selection across all simulation scenarios, consistent with existing cAIC approaches \citep{saefken2014unifying,wood2016smoothing}. Specifically, it yields a moderately higher null model rejection rate than the marginal AIC, and this rate converges to one as the strength of the alternative model increases, while remaining higher than that of the marginal AIC throughout. By contrast, \citet{yu2012conditional} exhibited anomalous behavior, with the null model rejection rate occasionally falling below that of the marginal AIC at higher levels of alternative hypothesis strength. In the case study presented in Sec.~\ref{sec:fish_stock_model_selection}, the cAIC computed using Eq.~(\ref{EQ:CAIC_October_18_2024}) selected a simpler model than that favored by the marginal AIC. Collectively, these results highlight the potential of Eq.~(\ref{EQ:CAIC_October_18_2024}) as a robust cAIC estimator for model selection across a wide range of applications.

For linear mixed-effects models, Eq. (\ref{EQ:CAIC_October_15_2024}) and \citet{overholser2014effective} address correlated errors. However, these correlations can instead be represented through correlated random effects (REs) combined with independent errors. Under this reparameterization, the cAIC can be computed using Eq. (\ref{EQ:CAIC_October_15_2024}) or \citet{liang2008note} without requiring the error-covariance matrix, thereby avoiding the bias introduced by estimating it.

Although the $O(1/n_t)$ term in (\ref{EQ:CAIC_October_18_2024}) is equal to $-2\mathrm{trace}\{\Sigma_u^{-1}(\partial\hat{\Psi}/\partial\theta^{\top})\mathrm{COV}(\hat{\theta})(\partial\hat{\Psi}^{\top}/\partial\theta)\}$ when $p(u)$ follows a multivariate normal distribution with covariance matrix $\Sigma_u$, our unpublished simulation results suggest that including this term may not improve the performance of the estimator in (\ref{EQ:CAIC_October_18_2024}). In particular, when the true value of the positive parameter $\delta$ in (\ref{EQ:RE_model_October_21_2024}) is 0, lying on the boundary of its parameter space, the variance of the maximum marginal likelihood estimator of $\delta$ cannot be estimated by the negative inverse of the Fisher information, making the $O(1/n_t)$ term difficult to evaluate. Further improvements to the method (\ref{EQ:CAIC_October_18_2024}) could focus on alternative approaches to evaluate or approximate the final term in the generic Equation (\ref{EQ:CAIC_exact_October_17_2024}). For certain models, such as those in the exponential family, this term may also have a closed-form expression, enabling more accurate cAIC calculations for these cases. 

The definition of cAI in Eq.~(\ref{EQ:cAI_October_9_2024}), and hence the final term in the cAIC expression in Eq.~(\ref{EQ:CAIC_exact_October_17_2024}), involves integration with respect to the true RE distribution $p(u)$. Consequently, unifying cAIC formulations, such as Eq.~(\ref{EQ:CAIC_October_18_2024}) and those of \citet{yu2012conditional} and \citet{wood2016smoothing}, explicitly depend on assumptions about the RE distribution. This dependence raises important questions regarding robustness to RE distribution misspecification and the ability of the proposed cAIC methods to correctly identify the true RE structure, particularly the covariance of a multivariate normal RE, among competing candidates. Similarly, the robustness of the proposed cAIC methods to likelihood ($l_c$ in Eq.~\ref{EQ:log_joint_likelihood_October_8_2024}) misspecification, such as an incorrect mean structure, and their ability to select the true model among multiple candidates, also warrant investigation. A comprehensive assessment of these issues is beyond the scope of the present work and is left for future research.

\FloatBarrier
\section*{Acknowledgments}
We sincerely thank the three reviewers and the associate editors for their insightful and constructive comments and suggestions, which substantially improved the quality and presentation of this paper.
\if0\blind
{
Research funding to NZ was provided by the Natural Sciences and Engineering Research Council of Canada [RGPIN-2024-04746].
Research funding to NC was provided by 1) the Ocean Choice
International Industry Research Chair program at the Marine Institute
of Memorial University of Newfoundland, 2) the Ocean Frontier Institute, through an award from the Canada First Research Excellence Fund, and 3) the Natural Sciences and Engineering Research Council of Canada [RGPIN-2016-04307].
} \fi

\bibliographystyle{Chicago}

\bibliography{mybib_marine}

@INPROCEEDINGS{Akaike1973AIC,
	author = {Akaike, Hirotsugu},
	title = {Information theory and an extension of the maximum likelihood principle},
	booktitle = {Second International Symposium on Information Theory},	
	year = 1973,	
	publisher = {Budapest: Akademiai Kiado},
	editor = {Petrov, B. and Csaki, F.},
	pages = {267–81},
}

@article{donohue2011conditional,
	title={Conditional Akaike information under generalized linear and proportional hazards mixed models},
	author={Donohue, MC and Overholser, R and Xu, R and Vaida, F},
	journal={Biometrika},
	volume={98},
	number={3},
	pages={685--700},
	year={2011},
	publisher={Oxford University Press}
}

@article{gudmundsson1994time,
	title={Time series analysis of catch-at-age observations},
	author={Gudmundsson, Gudmundur},
	journal={Journal of the Royal Statistical Society: Series C (Applied Statistics)},
	volume={43},
	number={1},
	pages={117--126},
	year={1994},
	publisher={Wiley Online Library}
}

@article{hooten2015guide,
	title={A guide to Bayesian model selection for ecologists},
	author={Hooten, Mevin B and Hobbs, N Thompson},
	journal={Ecological monographs},
	volume={85},
	number={1},
	pages={3--28},
	year={2015},
	publisher={Wiley Online Library}
}

@book{ingersoll1987theory,
	title={Theory of financial decision making},
	author={Ingersoll, Jonathan E},
	volume={3},
	year={1987},
	publisher={Rowman \& Littlefield}
}

@article{cadigan2015state,
	title={A state-space stock assessment model for northern cod, including under-reported catches and variable natural mortality rates},
	author={Cadigan, Noel G},
	journal={Canadian Journal of Fisheries and Aquatic Sciences},
	volume={73},
	number={2},
	pages={296--308},
	year={2015},
	publisher={NRC Research Press}
}

@article{cadigan2022complex,
	title={The complex relationship between weight and length of Atlantic cod off the south coast of Newfoundland},
	author={Cadigan, Noel and Robertson, Matthew D and Nirmalkanna, Kunasekaran and Zheng, Nan},
	journal={Canadian Journal of Fisheries and Aquatic Sciences},
	volume={79},
	number={11},
	pages={1798--1819},
	year={2022},
	publisher={Canadian Science Publishing 1840 Woodward Drive, Suite 1, Ottawa, ON K2C 0P7}
}

@article{kawakubo2014modified,
	title={Modified conditional AIC in linear mixed models},
	author={Kawakubo, Yuki and Kubokawa, Tatsuya},
	journal={Journal of Multivariate Analysis},
	volume={129},
	pages={44--56},
	year={2014},
	publisher={Elsevier}
}

@article{lindstrom1990nonlinear,
	title={Nonlinear mixed effects models for repeated measures data},
	author={Lindstrom, Mary J and Bates, Douglas M},
	journal={Biometrics},
	pages={673--687},
	year={1990},
	publisher={JSTOR}
}

@Article{lme4,
	title = {Fitting Linear Mixed-Effects Models Using {lme4}},
	author = {Douglas Bates and Martin M{\"a}chler and Ben Bolker and Steve Walker},
	journal = {Journal of Statistical Software},
	year = {2015},
	volume = {67},
	number = {1},
	pages = {1--48},
	doi = {10.18637/jss.v067.i01},
}

@article{maunder2011state,
	title={{A state--space multistage life cycle model to evaluate population impacts in the presence of density dependence: illustrated with application to delta smelt ({\it Hyposmesus transpacificus)}}},
	author={Maunder, Mark N and Deriso, Richard B},
	journal={Canadian Journal of Fisheries and Aquatic Sciences},
	volume={68},
	number={7},
	pages={1285--1306},
	year={2011},
	publisher={NRC Research Press}
}

@article{hudson1978natural,
	title={A natural identity for exponential families with applications in multiparameter estimation},
	author={Hudson, H Malcolm},
	journal={The Annals of Statistics},
	volume={6},
	number={3},
	pages={473--484},
	year={1978},
	publisher={Institute of Mathematical Statistics}
}

@book{jorgensen1997theory,
	title={The theory of dispersion models},
	author={Jorgensen, Bent},
	year={1997},
	publisher={Chapman and Hall, London}
}

@article{kristensen2016tmb,
	title={TMB: automatic differentiation and Laplace approximation},
	author={Kristensen, Kasper and Nielsen, Anders and Berg, Casper W and Skaug, Hans and Bell, Brad},
	journal={Journal of Statistical Software},
	volume={70},
	number={5},
	pages={1–21},
	year={2016}
}

@article{liang2008note,
	title={A note on conditional AIC for linear mixed-effects models},
	author={Liang, Hua and Wu, Hulin and Zou, Guohua},
	journal={Biometrika},
	volume={95},
	number={3},
	pages={773--778},
	year={2008},
	publisher={Oxford University Press}
}

@article{overholser2014effective,
	title={Effective degrees of freedom and its application to conditional AIC for linear mixed-effects models with correlated error structures},
	author={Overholser, Rosanna and Xu, Ronghui},
	journal={Journal of multivariate analysis},
	volume={132},
	pages={160--170},
	year={2014},
	publisher={Elsevier}
}

@Manual{r2018citation,
	title = {R: A Language and Environment for Statistical Computing},
	author = {{R Core Team}},
	organization = {R Foundation for Statistical Computing},
	address = {Vienna, Austria},
	year = {2022},
	url = {https://www.R-project.org/},
}

@article{saefken2014unifying,
	title={A unifying approach to the estimation of the conditional Akaike information in generalized linear mixed models},
	author={Saefken, Benjamin and Kneib, Thomas and van Waveren, Clara-Sophie and Greven, Sonja},
	year={2014}
}

@article{safken2021conditional,
	title={Conditional model selection in mixed-effects models with cAIC4},
	author={S{\"a}fken, Benjamin and R{\"u}gamer, David and Kneib, Thomas and Greven, Sonja},
	journal={Journal of Statistical Software},
	volume={99},
	pages={1--30},
	year={2021}
}

@inproceedings{stein1972bound,
	title={A bound for the error in the normal approximation to the distribution of a sum of dependent random variables},
	author={Stein, Charles},
	booktitle={Proceedings of the sixth Berkeley symposium on mathematical statistics and probability, volume 2: Probability theory},
	volume={6},
	pages={583--603},
	year={1972},
	organization={University of California Press}
}

@Article{stock2021implementing,
	author    = {Stock, Brian C and Xu, Haikun and Miller, Timothy J and Thorson, James T and Nye, Janet A},
	journal   = {Fisheries Research},
	title     = {Implementing two-dimensional autocorrelation in either survival or natural mortality improves a state-space assessment model for Southern New England-Mid Atlantic yellowtail flounder},
	year      = {2021},
	pages     = {105873},
	volume    = {237},
	publisher = {Elsevier},
}

@article{vaida2005conditional,
	title={Conditional Akaike information for mixed effects models},
	author={Vaida, Florin and Blanchard, Suzette},
	journal={Biometrika},
	volume={92},
	number={2},
	pages={351-370},
	year={2005}
}

@Article{pedersen2011estimation,
  author    = {Pedersen, Martin W{\ae}ver and Berg, Casper Willestofte and Thygesen, Uffe H{\o}gsbro and Nielsen, Anders and Madsen, Henrik},
  title     = {Estimation methods for nonlinear state-space models in ecology},
  journal   = {Ecological Modelling},
  year      = {2011},
  volume    = {222},
  number    = {8},
  pages     = {1394--1400},
  publisher = {Elsevier},
}

@Article{nielsen2014estimation,
  author    = {Nielsen, Anders and Berg, Casper W},
  title     = {Estimation of time-varying selectivity in stock assessments using state-space models},
  journal   = {Fisheries Research},
  year      = {2014},
  volume    = {158},
  pages     = {96--101},
  publisher = {Elsevier},
}

@Article{schnute1994general,
  author    = {Schnute, Jon T},
  title     = {A general framework for developing sequential fisheries models},
  journal   = {Canadian Journal of Fisheries and Aquatic Sciences},
  year      = {1994},
  volume    = {51},
  number    = {8},
  pages     = {1676--1688},
  publisher = {NRC Research Press},
}

@Article{aanes2007estimation,
  author    = {Aanes, Sondre and Engen, Steinar and S{\ae}ther, Bernt-Erik and Aanes, Ronny},
  title     = {Estimation of the parameters of fish stock dynamics from catch-at-age data and indices of abundance: can natural and fishing mortality be separated?},
  journal   = {Canadian Journal of fisheries and aquatic sciences},
  year      = {2007},
  volume    = {64},
  number    = {8},
  pages     = {1130--1142},
  publisher = {NRC Research Press},
}

@Article{zheng2021frequentist,
  author    = {Zheng, Nan and Cadigan, Noel},
  journal   = {Computational Statistics \& Data Analysis},
  title     = {Frequentist delta-variance approximations with mixed-effects models and TMB},
  year      = {2021},
  pages     = {107227},
  volume    = {160},
  publisher = {Elsevier},
}

@Article{kass1989approximate,
  author    = {Kass, Robert E and Steffey, Duane},
  title     = {Approximate Bayesian inference in conditionally independent hierarchical models (parametric empirical Bayes models)},
  journal   = {Journal of the American Statistical Association},
  year      = {1989},
  volume    = {84},
  number    = {407},
  pages     = {717--726},
  publisher = {Taylor \& Francis Group},
}

@Article{perreault2020state,
  author  = {Perreault, Andrea MJ and Wheeland, Laura J and Morgan, M Joanne and Cadigan, Noel G},
  journal = {J. Northw. Atl. Fish. Sci},
  title   = {A state-space stock assessment model for American plaice on the Grand Bank of Newfoundland},
  year    = {2020},
  pages   = {45--104},
  volume  = {51},
}

@Article{pedersen2017stochastic,
  author    = {Pedersen, Martin W and Berg, Casper W},
  journal   = {Fish and Fisheries},
  title     = {A stochastic surplus production model in continuous time},
  year      = {2017},
  number    = {2},
  pages     = {226--243},
  volume    = {18},
  publisher = {Wiley Online Library},
}

@article{wood2016smoothing,
	title={Smoothing parameter and model selection for general smooth models},
	author={Wood, Simon N and Pya, Natalya and S{\"a}fken, Benjamin},
	journal={Journal of the American Statistical Association},
	volume={111},
	number={516},
	pages={1548--1563},
	year={2016},
	publisher={Taylor \& Francis}
}

@article{yu2012conditional,
	title={Conditional Akaike information criterion for generalized linear mixed models},
	author={Yu, Dalei and Yau, Kelvin KW},
	journal={Computational Statistics \& Data Analysis},
	volume={56},
	number={3},
	pages={629--644},
	year={2012},
	publisher={Elsevier}
}

@article{Zheng2021P2,
	title={Frequentist Conditional Variance for Nonlinear Mixed-Effects Models},
	author={Zheng, Nan and Cadigan, Noel},
	journal={Journal of Statistical Theory and Practice},
	volume={17},
	number={1},
	pages={1--30},
	year={2023},
	publisher={Springer}
}

@Article{stock2021woods,
  author    = {Stock, Brian C and Miller, Timothy J},
  journal   = {Fisheries Research},
  title     = {The Woods Hole Assessment Model (WHAM): A general state-space assessment framework that incorporates time-and age-varying processes via random effects and links to environmental covariates},
  year      = {2021},
  pages     = {105967},
  volume    = {240},
  publisher = {Elsevier},
}

\begin{appendices}
	\section{Derivation of Eqs. (\ref{EQ:derivatives_hattheta_Psi_wrt_y_October_10_2024})}	\label{app:derivatives_derivation_October_30_2024}
	Assume that the log-marginal likelihood $l(y,\theta)$ and log-joint likelihood $l_j(y,\theta,\Psi)$ have well-defined second-order derivatives. The maximum marginal likelihood estimator of model parameters, $\hat{\theta}$, is defined by
	\begin{linenomath*} 
		\begin{align*}
			\dfrac{\partial l(y,\hat{\theta})}{\partial\hat{\theta}} &\equiv 0,
		\end{align*}
	\end{linenomath*} 	
	which is true for any data $y$. Therefore,
	\begin{linenomath*} 
		\begin{align}\label{dtheta_dy_July_21_2024}
			\begin{split}
				&\dfrac{d}{dy}\dfrac{\partial l(y,\hat{\theta})}{\partial\hat{\theta}^{\top}} = 0 \Rightarrow
				\dfrac{\partial^2 l(y,\hat{\theta})}{\partial y\partial\hat{\theta}^{\top}} + \dfrac{d\hat{\theta}^{\top}}{d y}\dfrac{\partial^2 l(y,\hat{\theta})}{\partial\hat{\theta}\partial\hat{\theta}^{\top}} =0 \\
				&\Rightarrow \dfrac{d\hat{\theta}^{\top}}{d y} = -\dfrac{\partial^2 l(y,\hat{\theta})}{\partial y\partial\hat{\theta}^{\top}}\left\lbrace \dfrac{\partial^2 l(y,\hat{\theta})}{\partial\hat{\theta}\partial\hat{\theta}^{\top}} \right\rbrace^{-1}. 
			\end{split}			 
		\end{align}	
	\end{linenomath*} 	
	Similarly, the empirical Bayes RE predictor, $\hat{\Psi}(\hat{\theta})$, is defined by
	\begin{linenomath*} 
		\begin{align*}
			\dfrac{\partial l_j(y,\hat{\theta},\hat{\Psi})}{\partial\hat{\Psi}} &\equiv 0,
		\end{align*}
	\end{linenomath*} 	
	which is true for any $y$ and $\hat{\theta}$. Therefore,
	\begin{linenomath*} 
		\begin{align}\label{dPsi_dy_July_21_2024}
			\begin{split}
				\dfrac{d}{dy} \dfrac{\partial l_j(y,\hat{\theta},\hat{\Psi})}{\partial\hat{\Psi}^{\top}} &= 0 = \dfrac{\partial^2 l_j(y,\hat{\theta},\hat{\Psi})}{\partial y\partial\hat{\Psi}^{\top}} + \dfrac{d\hat{\theta}^{\top}}{d y}\dfrac{\partial^2 l_j(y,\hat{\theta},\hat{\Psi})}{\partial \hat{\theta}\partial\hat{\Psi}^{\top}} + \dfrac{d\hat{\Psi}^{\top}}{d y}\dfrac{\partial^2 l_j(y,\hat{\theta},\hat{\Psi})}{\partial \hat{\Psi}\partial\hat{\Psi}^{\top}} \Rightarrow\\
				\dfrac{d\hat{\Psi}^{\top}}{d y} &= -\left\lbrace \dfrac{\partial^2 l_j(y,\hat{\theta},\hat{\Psi})}{\partial y\partial\hat{\Psi}^{\top}} + \dfrac{d\hat{\theta}^{\top}}{d y}\dfrac{\partial^2 l_j(y,\hat{\theta},\hat{\Psi})}{\partial \hat{\theta}\partial\hat{\Psi}^{\top}} \right\rbrace \left\lbrace \dfrac{\partial^2 l_j(y,\hat{\theta},\hat{\Psi})}{\partial \hat{\Psi}\partial\hat{\Psi}^{\top}} \right\rbrace^{-1},
			\end{split}
		\end{align}
	\end{linenomath*} 	
	where $d\hat{\theta}^{\top}/dy$ is given by (\ref{dtheta_dy_July_21_2024}).
	
	\section{Derivation of Eq. (\ref{EQ:CAIC_October_15_2024})}\label{app:derivation_4_October_30_2024}	
	According to \cite{vaida2005conditional}, the conditional Akaike information (cAI) is defined with a new set of data $y^*$ generated independently from $g(y^*|u)$ as
	\begin{linenomath*} 
		\begin{align}\label{cAI_defining_September_14_2024}
			\mbox{cAI} = &-2\,\mathrm{E}_{g(y,u)}\mathrm{E}_{g(y^*|u)}l_c\{\hat{\theta}(y),\hat{\Psi}(y)|y^*\}\\
			= &-2\,\mathrm{E}_{g(y,u)}\mathrm{E}_{g(y^*|u)}\left[ l_c(\theta_o,\Psi_o|y^*) + \dfrac{\partial l_c(\Theta_o|y^*)}{\partial\Theta_o^{\top}}\{\hat{\Theta}(y) - \Theta_o\} \right. \\
			&\left. + \dfrac{1}{2}\{\hat{\Theta}(y) - \Theta_o\}^{\top}\dfrac{\partial^2 l_c(\Theta_o|y^*)}{\partial\Theta_o\partial\Theta_o^{\top}}\{\hat{\Theta}(y) - \Theta_o\} \right] \nonumber\\ 
			= &-2\,\mathrm{E}_{g(y,u)}\left[ \vphantom{\dfrac{\partial^2 l_c(\Theta_o|y^*)}{\partial\Theta_o\partial\Theta_o^{\top}}} \mathrm{E}_{g(y^*|u)}\left\lbrace l_c(\theta_o,\Psi_o|y^*)\right\rbrace \right. \\
			&\left. +  \dfrac{1}{2}\{\hat{\Theta}(y) - \Theta_o\}^{\top}\mathrm{E}_{g(y^*|u)}\left\lbrace\dfrac{\partial^2 l_c(\Theta_o|y^*)}{\partial\Theta_o\partial\Theta_o^{\top}}\right\rbrace\{\hat{\Theta}(y) - \Theta_o\} \right], \nonumber
		\end{align}
	\end{linenomath*} 	
	where $\hat{\Theta}(y)$ is a function of $y$ which is independent of $y^*$. The observed $l_c$ is
	\begin{linenomath*} 
		\begin{align}\label{lc_August_4_2024}
			\begin{split}
				l_c(\hat{\theta},\hat{\Psi}|y) &= l_c(\Theta_o|y) + (\hat{\Theta} - \Theta_o)^{\top}\dfrac{\partial l_c(\Theta_o|y)}{\partial\Theta_o}+ \dfrac{1}{2}(\hat{\Theta} - \Theta_o)^{\top}\dfrac{\partial^2 l_c(\Theta_o|y)}{\partial\Theta_o\partial\Theta_o^{\top}}(\hat{\Theta} - \Theta_o).
			\end{split}
		\end{align}
	\end{linenomath*} 	
	Let $\mathrm{E}_{g(y|u)}(y)=\mu_y$, and $\mathrm{E}_{g(y,u)}(\hat{\Theta}-\Theta_o)\approx0$ according to \cite{zheng2021frequentist}. By denoting the marginal expectation $\mathrm{E}_{g(y,u)}(\cdot)$ as $\mathrm{E}(\cdot)$, We evaluate
	\begin{linenomath*} 
		\begin{align}\label{expectation_lc_September_14_2024}
			\begin{split}
				-2\mathrm{E}\{l_c(\hat{\theta},\hat{\Psi}|y)\} &= -2\mathrm{E}\left\lbrace l_c(\Theta_o|y) + (\hat{\Theta} - \Theta_o)^{\top}\dfrac{\partial l_c(\Theta_o|y)}{\partial\Theta_o}\right. \\
				&\left.\hspace{4.5mm} + \dfrac{1}{2}(\hat{\Theta} - \Theta_o)^{\top}\dfrac{\partial^2 l_c(\Theta_o|y)}{\partial\Theta_o\partial\Theta_o^{\top}}(\hat{\Theta} - \Theta_o)\right\rbrace \\
				&= -2\mathrm{E}\left\lbrace l_c(\Theta_o|y) + (y-\mu_y)^{\top}\dfrac{\partial\hat{\Theta}^{\top}(\mu_y)}{\partial \mu_y}\left. \dfrac{\partial^2 l_c(\Theta_o|y)}{\partial\Theta_o\partial y^{\top}}\right|_{y=\mu_y}(y-\mu_y) \right. \\
				&\left.\hspace{4.5mm} + \dfrac{1}{2}(\hat{\Theta} - \Theta_o)^{\top}\dfrac{\partial^2 l_c(\Theta_o|y)}{\partial\Theta_o\partial\Theta_o^{\top}}(\hat{\Theta} - \Theta_o)\right\rbrace.
			\end{split}
		\end{align}
	\end{linenomath*} 	
	If we neglect the difference between the second order terms of (\ref{cAI_defining_September_14_2024}) and (\ref{expectation_lc_September_14_2024}) (our unpublished numerical study also indicates that the difference is small), the bias from using $-2l_c(\hat{\theta},\hat{\Psi}|y)$ to estimate cAI formulated by (\ref{cAI_defining_September_14_2024}) is
	\begin{linenomath*} 
		\begin{align*}%\label{lc_August_4_2024}
			\begin{split}
				-2\, \mathrm{trace}\left\lbrace \dfrac{\partial\hat{\Theta}^{\top}(y)}{\partial y}\left. \dfrac{\partial^2 l_c(\Theta_o|y)}{\partial\Theta_o\partial y^{\top}}\right|_{y=\mu_y}\mathrm{Cov}(y) \right\rbrace .
			\end{split}
		\end{align*}
	\end{linenomath*} 		
	
	\section{Derivation of Eq. (\ref{EQ:CAIC_October_18_2024})}\label{app:derivation_6_October_30_2024}
	We continue with Eqs. (\ref{cAI_defining_September_14_2024}) and (\ref{lc_August_4_2024}), which hold for any set of observations.
	
	For clarity, we denote the fixed effects (parameters) in $l_c$ as $\phi$. Then the log-marginal distribution of the REs $\Psi$ is written as $l_r(\tau,\Psi)$, where $\tau$ represents the associated fixed parameters. By assumption, $f(y|\phi_o,\Psi_o)=g(y|u)$. We examine the $i$th component of $\phi$ and note that the estimator $\hat{\Theta}$ depends solely on $y$, with no dependence on $\Theta_o$,
	\begin{linenomath*} 
		\begin{align}
			\begin{split}
				&\mathrm{E}_{p(u)}\mathrm{E}_{g(y|u)}\left\lbrace (\hat{\phi}_i-\phi_{o,i})\dfrac{\partial l_c(\phi_o,\Psi_o|y)}{\partial\phi_{o,i}}\right\rbrace \\
				&= \mathrm{E}_{p(u)}\int (\hat{\phi}_i-\phi_{o,i})\dfrac{\partial f(y|\phi_o,\Psi_o)/\partial\phi_{o,i}}{f(y|\phi_o,\Psi_o)}g(y|u) dy\\
				&= \mathrm{E}_{p(u)}\int (\hat{\phi}_i-\phi_{o,i})\dfrac{\partial f(y|\phi_o,\Psi_o)}{\partial\phi_{o,i}} dy\\
				&= \mathrm{E}_{p(u)}\left\lbrace \dfrac{\partial }{\partial\phi_{o,i}}\int (\hat{\phi}_i-\phi_{o,i})f(y|\phi_o,\Psi_o) dy - \int \dfrac{\partial(\hat{\phi}_i-\phi_{o,i})}{\partial\phi_{o,i}}f(y|\phi_o,\Psi_o) dy \right\rbrace \\
				&= \dfrac{\partial }{\partial\phi_{o,i}}\mathrm{E}_{p(u)}\mathrm{E}_{g(y|u)}(\hat{\phi}_i-\phi_{o,i}) + 1 = 1.
			\end{split}
		\end{align}
	\end{linenomath*}	
	Here we can exchange $\mathrm{E}_{p(u)}$ and $\partial/\partial\phi_{o,i}$ because $p(u)$ does not depend on $\phi_o$. We also applied $\mathrm{E}(\hat{\phi}_i-\phi_{o,i})=0$. Therefore, 
	\begin{linenomath*} 
		\begin{align}
			\begin{split}
				\mathrm{E}_{p(u)}\mathrm{E}_{g(y|u)}\left\lbrace (\hat{\phi}-\phi_o)^{\top}\dfrac{\partial l_c(\phi_o,\Psi_o|y)}{\partial\phi_o}\right\rbrace &= p_c,
			\end{split}
		\end{align}
	\end{linenomath*}	
	where $p_c$ is the number of parameters in $l_c(\phi_o,\Psi_o|y)$, or equivalently the length of $\phi_o$.
	
	Assume that $\Psi_o$ is modeled with a multivariate normal distribution with mean $\mu_{\Psi}$ and covariance $\Sigma$. Then, according to the the conditional inference outlined in \cite{Zheng2021P2},
	\begin{linenomath*} 
		\begin{align}\label{EQ:Psi_1st_order_term_November_2_2024}
			\begin{split}
				&\mathrm{E}_{p(u)}\mathrm{E}_{g(y|u)}\left\lbrace (\hat{\Psi}_i-\Psi_{o,i})\dfrac{\partial l_c(\phi_o,\Psi_o|y)}{\partial\Psi_{o,i}}\right\rbrace \\
				&= \mathrm{E}_{p(u)}\int (\hat{\Psi}_i-\Psi_{o,i})\dfrac{\partial f(y|\phi_o,\Psi_o)/\partial\Psi_{o,i}}{f(y|\phi_o,\Psi_o)}g(y|u) dy\\
				&= \mathrm{E}_{p(u)}\int (\hat{\Psi}_i-\Psi_{o,i})\dfrac{\partial f(y|\phi_o,\Psi_o)}{\partial\Psi_{o,i}} dy\\
				&= \mathrm{E}_{p(u)}\left\lbrace \dfrac{\partial }{\partial\Psi_{o,i}}\int (\hat{\Psi}_i-\Psi_{o,i})f(y|\phi_o,\Psi_o) dy - \int \dfrac{\partial(\hat{\Psi}_i-\Psi_{o,i})}{\partial\Psi_{o,i}}f(y|\phi_o,\Psi_o) dy \right\rbrace \\
				&= \mathrm{E}_{p(u)}\dfrac{\partial }{\partial\Psi_{o,i}}\mathrm{E}_{g(y|u)}(\hat{\Psi}_i-\Psi_{o,i}) + 1\\
				&=-\mathrm{E}_{p(u)}\dfrac{\partial }{\partial\Psi_{o,i}}\mathbf{0}_i^{\top}\ddot{l}^{-1}_j(\phi_o,\tau_o,\Psi_o|y)\ddot{l}_r(\tau_o,\Psi_o)(\Psi_o-\mu_{\Psi}) + 1 \\
				&\hspace{4.5mm}+ \mathrm{E}_{p(u)}\dfrac{\partial }{\partial\Psi_{o,i}}\mathrm{E}_{g(y|u)}(\hat{\theta}-\theta_o)^{\top}\dfrac{\partial\hat{\Psi}_i}{\partial\theta_o}\\
				&=-\mathrm{E}_{p(u)}\mathbf{0}_i^{\top}\ddot{l}^{-1}_j(\phi_o,\tau_o,\Psi_o|y)\ddot{l}_r(\tau_o,\Psi_o)\mathbf{0}_i + 1 + \mathbf{0}_i^{\top} \mathrm{E}_{p(u)}\dfrac{\partial }{\partial\Psi_o}\mathrm{E}_{g(y|u)}(\hat{\theta}-\theta_o)^{\top}\dfrac{\partial\hat{\Psi}^{\top}}{\partial\theta_o} \mathbf{0}_i,
			\end{split}
		\end{align}
	\end{linenomath*}	
	where the double dots denotes hessian with respect to $\Psi_o$, and $\mathbf{0}_i$ is a vector of 0's with the $i$th element being 1. Therefore, 
	\begin{linenomath*} 
		\begin{align}
			\begin{split}
				\mathrm{E}_{p(u)}\mathrm{E}_{g(y|u)}\left\lbrace (\hat{\Psi}-\Psi_o)^{\top}\dfrac{\partial l_c(\phi_o,\Psi_o|y)}{\partial\Psi_o}\right\rbrace = &- \mbox{trace}\left\lbrace\ddot{l}^{-1}_j(\phi_o,\tau_o,\Psi_o|y) \ddot{l}_r(\tau_o,\Psi_o)\right\rbrace + q \\
				&\hspace{-25.5mm}+ \mbox{trace}\left\lbrace \mathrm{E}_{p(u)}\dfrac{\partial }{\partial\Psi_o}\mathrm{E}_{g(y|u)}(\hat{\theta}-\theta_o)^{\top}\dfrac{\partial\hat{\Psi}^{\top}}{\partial\theta_o} \right\rbrace\\
				&\hspace{-30mm}= - \mbox{trace}\left\lbrace\ddot{l}^{-1}_j(\phi_o,\tau_o,\Psi_o|y) \ddot{l}_r(\tau_o,\Psi_o)\right\rbrace + q + O(1/n_t),
			\end{split}
		\end{align}
	\end{linenomath*}	
	where $q$ is the number of random effects, and we prove the $O(1/n_t)$ order in the Supplementary Material.
	
	Thus,
	\begin{linenomath*} 
		\begin{align}\label{Elc_August_4_2024}
			\begin{split}
				\mathrm{E}\left\lbrace l_c(\hat{\phi},\hat{\Psi}|y)\right\rbrace  &= \mathrm{E}\left\lbrace l_c(\Theta_o|y)\right\rbrace  + \mathrm{E}\left\lbrace (\hat{\Theta} - \Theta_o)^{\top}\dfrac{\partial l_c(\Theta_o|y)}{\partial\Theta_o}\right\rbrace\\ &\hspace{4.5mm}+\mathrm{E}\left\lbrace  \dfrac{1}{2}(\hat{\Theta} - \Theta_o)^{\top}\dfrac{\partial^2 l_c(\Theta_o|y)}{\partial\Theta_o\partial\Theta_o^{\top}}(\hat{\Theta} - \Theta_o)\right\rbrace \\		
				&= \mathrm{E}\left\lbrace l_c(\Theta_o|y)\right\rbrace + p_c + q - \mbox{trace}\left\lbrace \ddot{l}^{-1}_j(\phi_o,\tau_o,\Psi_o|y)\ddot{l}_r(\tau_o,\Psi_o)\right\rbrace \\
				&\hspace{4.5mm}+ \mathrm{E}\left\lbrace  \dfrac{1}{2}(\hat{\Theta} - \Theta_o)^{\top}\dfrac{\partial^2 l_c(\Theta_o|y)}{\partial\Theta_o\partial\Theta_o^{\top}}(\hat{\Theta} - \Theta_o)\right\rbrace + O(1/n_t).
			\end{split}
		\end{align}
	\end{linenomath*} 	
	The bias for using $-2\,l_c(\hat{\phi},\hat{\Psi}|y)$ to estimate cAI is then
	\begin{linenomath*} 
		\begin{align}\label{Elc_August_4_2024}
			\begin{split}
				&-2\,\mathrm{E}\left\lbrace l_c(\hat{\phi},\hat{\Psi}|y)\right\rbrace - \mathrm{cAI} = -2\,\mathrm{E}\left\lbrace l_c(\Theta_o|y)\right\rbrace -2\, p_c -2\,q  + O(1/n_t) \\
				&+2\, \mbox{trace}\left\lbrace \ddot{l}^{-1}_j(\phi_o,\tau_o,\Psi_o|y)\ddot{l}_r(\tau_o,\Psi_o)\right\rbrace -\mathrm{E}\left\lbrace (\hat{\Theta} - \Theta_o)^{\top}\dfrac{\partial^2 l_c(\Theta_o|y)}{\partial\Theta_o\partial\Theta_o^{\top}}(\hat{\Theta} - \Theta_o)\right\rbrace\\
				& +2\,\mathrm{E}_{g(y,u)}\left[ \mathrm{E}_{g(y^*|u)}\left\lbrace l_c(\phi_o,\Psi_o|y^*)\right\rbrace + \dfrac{1}{2}(\hat{\Theta} - \Theta_o)^{\top}\mathrm{E}_{g(y^*|u)}\left\lbrace\dfrac{\partial^2 l_c(\Theta_o|y^*)}{\partial\Theta_o\partial\Theta_o^{\top}}\right\rbrace(\hat{\Theta} - \Theta_o) \right]\\
				&= -2\, p_c -2\,q  +2\, \mbox{trace}\left\lbrace\ddot{l}^{-1}_j(\phi_o,\tau_o,\Psi_o|y) \ddot{l}_r(\tau_o,\Psi_o)\right\rbrace + O(1/n_t).
			\end{split}
		\end{align}
	\end{linenomath*} 		
	Here we neglect the difference between the second order terms.	
	
\end{appendices}

\bigskip
\newpage
\sectionfont{\centering}
\section*{SUPPLEMENTARY MATERIAL}
%\begin{center}
%	{\large\bf SUPPLEMENTARY MATERIAL}
%\end{center}
\setcounter{figure}{0}
\setcounter{table}{0}
\setcounter{equation}{0}
\renewcommand{\thetable}{S\arabic{table}}
\renewcommand{\thefigure}{S\arabic{figure}} 
\renewcommand{\theequation}{S\arabic{equation}}
%\begin{description}
	
	\subsection*{Derivation of the $O(1/n_t)$ approximation order}   %Brief description. (file type)
	Assume that $p(u)$ is multivariate normal with mean $\mu_u$ and covariance matrix $\Sigma_u$. According to Stein's lemma \citep{ingersoll1987theory},
	\begin{linenomath*} 
		\begin{align}\label{EQ:Ont_term_November_2_2024}
			\begin{split}
				\mathrm{E}_{p(u)}\dfrac{\partial }{\partial\Psi_o}\mathrm{E}_{g(y|u)}(\hat{\theta}-\theta_o)^{\top}\dfrac{\partial\hat{\Psi}_i}{\partial\theta_o} &= \Sigma_u^{-1}\mathrm{E}_{g(y,u)}\left\lbrace (\Psi_o-\mu_u)(\hat{\theta}-\theta_o)^{\top}\right\rbrace \dfrac{\partial\hat{\Psi}_i}{\partial\theta_o}\\
				&= -\Sigma_u^{-1} \dfrac{\partial\hat{\Psi}}{\partial\theta_o^{\top}} \mathrm{COV}(\hat{\theta}) \dfrac{\partial\hat{\Psi}_i}{\partial\theta_o},
			\end{split}
		\end{align}
	\end{linenomath*}	
	according to Eq. (30) in \cite{zheng2021frequentist}. Thus, by Eq. (\ref{EQ:Psi_1st_order_term_November_2_2024}), the $O(1/n_t)$ is equal to 
	\begin{linenomath*} 
		\begin{align}\label{EQ:Ont_term_November_2_2024}
			\begin{split}
				\mathrm{trace}\left\lbrace  \mathrm{E}_{p(u)}\dfrac{\partial }{\partial\Psi_o}\mathrm{E}_{g(y|u)}(\hat{\theta}-\theta_o)^{\top}\dfrac{\partial\hat{\Psi}^{\top}}{\partial\theta_o} \right\rbrace &= -\mathrm{trace}\left\lbrace  \Sigma_u^{-1} \dfrac{\partial\hat{\Psi}}{\partial\theta_o^{\top}} \mathrm{COV}(\hat{\theta}) \dfrac{\partial\hat{\Psi}^{\top}}{\partial\theta_o} \right\rbrace.
			\end{split}
		\end{align}
	\end{linenomath*}	
	
	\subsection*{A gamma observation example with more complicated link function}
	We assume that, conditional on all the random effects (REs), the observations follow independent gamma distributions.
	For the gamma shape parameter, we examine a value of 3 for the skewed case and a value of 5 for the near-Gaussian case. We model the gamma scale parameter $t$ and age $a$, $\mbox{Scale}_{t,a}$, using a shifted and scaled logit transformation,
	\begin{linenomath*} 
		\begin{align*}%\label{EQ:gamma_shape_October_27_2024}
			\begin{split}
				\mbox{Scale}_{t,a} &= 0.5 + \dfrac{1.5\exp(\eta_{t,a})}{1+\exp(\eta_{t,a})}.
			\end{split}
		\end{align*}
	\end{linenomath*}  
	This parameter structure constrains the standard deviation of the observations to fall between 0.87 and 6.32. The total number of years is $T=50$. When applying Eq. (\ref{EQ:CAIC_October_18_2024}), we set $n_{\mathrm{out}}=1500$ and $n_{\mathrm{inner}}=15000$. For Eq. (\ref{EQ:CAIC_October_15_2024}), due to the higher computational cost of evaluating the derivatives in Eq. (\ref{EQ:derivatives_hattheta_Psi_wrt_y_October_10_2024}), we used $n_{\mathrm{out}}=1000$ and $n_{\mathrm{inner}}=10000$.
	
	The simulation results are shown in Table \ref{tab:Gamma_simulation_logit_October_28_2024}. Method (\ref{EQ:CAIC_October_15_2024}) performed well with small RBs in the near-Gaussian case (shape = 5), but failed with large RBs in the skewed case (shape = 3), where higher-order derivatives with respect to $y$ may be required. In contrast, method (\ref{EQ:CAIC_October_18_2024}) performed reliably across all levels of skewness in the observational distribution, showing small RBs across all simulation scenarios. $\rm{BC}_{\rm Yu}$ showed acceptable performance but was slightly inferior to Eq.~(\ref{EQ:CAIC_October_18_2024}), whereas $\rm{BC}_{\rm GAM}$ failed in most scenarios. 
	
	\begin{table}[ht]
		\centering
		\caption{Relative biases (RBs) in estimating bias correction for the conditional Akaike information, based on data simulated with the linear predictor (\ref{EQ:RE_model_October_21_2024}) and gamma observational distribution with shape parameter modeled via the shifted and scaled logit transformation. $\rm{BC}_{\rm Yu}$ denotes the BC estimator proposed by \citet{yu2012conditional}, and $\rm{BC}_{\rm GAM}$ denotes the BC estimator given in Eq. (\ref{EQ:BC_GAM_December_10_2025}).}
		\label{tab:Gamma_simulation_logit_October_28_2024}
		\hspace*{-2.3cm}
		\begin{tabular}{ccccccc|ccccccc}
			\toprule
			\multicolumn{3}{c}{Design parameters} &
			\multicolumn{4}{c|}{Relative biases} &
			\multicolumn{3}{c}{Design parameters} &
			\multicolumn{4}{c}{Relative biases} \\
			\cmidrule(lr){1-3}\cmidrule(lr){4-7}
			\cmidrule(lr){8-10}\cmidrule(lr){11-14}
			$n_{t,a}$ & Shape & $\delta$ & Eq.~(\ref{EQ:CAIC_October_18_2024}) & Eq.~(\ref{EQ:CAIC_October_15_2024}) & $\rm{BC}_{\rm Yu}$ &
			$\rm{BC}_{\rm GAM}$ & $n_{t,a}$ & Shape & $\delta$ & Eq.~(\ref{EQ:CAIC_October_18_2024}) & Eq.~(\ref{EQ:CAIC_October_15_2024}) & $\rm{BC}_{\rm Yu}$ &
			$\rm{BC}_{\rm GAM}$ \\ 
			\midrule
			3 & 3 & 0.01 & 0.019 & 0.260 & -0.056 & -0.479 & 5 & 3 & 0.01 & 0.055 & 0.322 & -0.070 & -0.883 \\ 
			3 & 3 & 0.10 & 0.010 & 0.290 & -0.071 & -1.839 & 5 & 3 & 0.10 & 0.065 & 0.197 & -0.040 & -2.125 \\ 
			3 & 3 & 0.40 & 0.023 & 0.245 & -0.033 & -2.852 & 5 & 3 & 0.40 & 0.016 & 0.161 & -0.049 & -1.650 \\ 
			3 & 3 & 0.80 & 0.018 & 0.167 & -0.041 & -0.725 & 5 & 3 & 0.80 & 0.037 & 0.171 & 0.006 & -0.105 \\ 
			3 & 5 & 0.01 & 0.028 & 0.047 & -0.016 & 0.567 & 5 & 5 & 0.01 & 0.042 & 0.029 & -0.024 & -2.174 \\ 
			3 & 5 & 0.10 & 0.043 & 0.087 & -0.041 & -2.240 & 5 & 5 & 0.10 & 0.032 & 0.045 & -0.081 & -1.388 \\ 
			3 & 5 & 0.40 & 0.049 & 0.058 & -0.056 & -1.846 & 5 & 5 & 0.40 & 0.016 & 0.056 & -0.027 & -0.525 \\ 
			3 & 5 & 0.80 & 0.046 & 0.064 & -0.012 & -0.068 & 5 & 5 & 0.80 & 0.058 & 0.025 & 0.016 & 0.012 \\ 
			\bottomrule
		\end{tabular}
	\end{table}	

\FloatBarrier

	\subsection*{A gamma observation example with more parameters in RE distribution}
	For the Gamma observations, we adopt the linear predictor in Eq.~(\ref{EQ:RE_model_more_RE_parameters_December_12_2025}) and continue to use the log-link function in Eq.~(\ref{EQ:log_link_NB_October_28_2024}). All other aspects of the simulation design follow those described in Sec.~\ref{sec:gamma_October_21_2024}. The simulation results are summarized in Table~\ref{tab:Gamma_simulation_more_RE_parameters_December_7_2025}.
	\begin{table}[ht]
		\centering
		\caption{Relative biases (RBs) in estimating bias correction for the conditional Akaike information, based on data simulated with the linear predictor (\ref{EQ:RE_model_more_RE_parameters_December_12_2025}) and gamma observational distribution with log-link function (\ref{EQ:log_link_NB_October_28_2024}). $\rm{BC}_{\rm Yu}$ denotes the BC estimator proposed by \citet{yu2012conditional}, and $\rm{BC}_{\rm GAM}$ denotes the BC estimator given in Eq.~(\ref{EQ:BC_GAM_December_10_2025}).}
		\label{tab:Gamma_simulation_more_RE_parameters_December_7_2025}	\hspace*{-0.7cm}	
		\begin{tabular}{cccccc|cccccc}
			\toprule
			\multicolumn{3}{c}{Design parameters} &
			\multicolumn{3}{c|}{Relative biases} &
			\multicolumn{3}{c}{Design parameters} &
			\multicolumn{3}{c}{Relative biases} \\
			\cmidrule(lr){1-3}\cmidrule(lr){4-6}
			\cmidrule(lr){7-9}\cmidrule(lr){10-12}
			$n_{t,a}$ & Shape & $\delta$ & Eq.~(\ref{EQ:CAIC_October_18_2024}) & $\rm{BC}_{\rm Yu}$ &
			$\rm{BC}_{\rm GAM}$ & $n_{t,a}$ & Shape & $\delta$ & Eq.~(\ref{EQ:CAIC_October_18_2024}) & $\rm{BC}_{\rm Yu}$ &
			$\rm{BC}_{\rm GAM}$ \\ 
			\midrule
			3 & 3 & 0.01 & -0.038 & -0.485 & -0.044   & 5 & 3 & 0.01 & -0.012 & -0.451 & -0.016 \\
			3 & 3 & 0.10 & -0.036 & -0.221 & -0.038   & 5 & 3 & 0.10 & -0.008 & -0.092 & -0.009 \\
			3 & 3 & 0.40 & -0.059 & -0.080 & -0.074   & 5 & 3 & 0.40 & -0.041 & -0.077 & -0.057 \\
			3 & 3 & 0.80 & -0.082 & -0.100 & -0.099   & 5 & 3 & 0.80 & -0.044 & -0.059 & -0.062 \\
			3 & 5 & 0.01 & -0.028 & -0.483 & -0.035   & 5 & 5 & 0.01 & -0.019 & -0.471 & -0.025 \\
			3 & 5 & 0.10 & -0.031 & -0.126 & -0.034   & 5 & 5 & 0.10 & -0.021 & -0.051 & -0.025 \\
			3 & 5 & 0.40 & -0.036 & -0.110 & -0.053   & 5 & 5 & 0.40 & -0.025 & -0.038 & -0.042 \\
			3 & 5 & 0.80 & -0.050 & -0.070 & -0.068   & 5 & 5 & 0.80 & -0.019 & -0.038 & -0.038 \\
			\bottomrule
		\end{tabular}
	\end{table}
	\FloatBarrier
	
	\clearpage
	\subsection*{R and TMB Codes for the Gaussian simulation example in Sec. \ref{sec:Gaussian_October_21_2024}} The codes include simulation\_Gaussian\_example.R, fit.cpp, fit\_dataparameters.cpp, and fit\_sim.cpp. simulation\_Gaussian\_example.R includes simulation setup, and the function cAIC\_calc() that implements Eqs. (\ref{EQ:CAIC_October_15_2024}) and (\ref{EQ:CAIC_October_18_2024}) to evaluate the bias correction (BC) term in cAIC. The outputs of this function are cAIC using (\ref{EQ:CAIC_October_18_2024}), BC using (\ref{EQ:CAIC_October_15_2024}) and BC using (\ref{EQ:CAIC_October_18_2024}). fit.cpp serves as the main code for model fitting and data generation. fit\_dataparameters.cpp is a revision to fit.cpp by declaring data as parameters for implementing (\ref{EQ:CAIC_October_15_2024}). fit\_sim.cpp evaluate the integration over $y^*$ in Eq. (\ref{EQ:cAI_October_9_2024}) by generating and taking average over $y^*$ for evaluating the true BC.
	
	

\section*{Supplementary material}

```r
# simulation_Gaussian_example.R

library(TMB)
library(ggplot2)
library(patchwork)
library(xtable)
library(randtests)
library(lmtest)
library(copula)
library(EnvStats)

library(Matrix)
library(expm)
library(matrixcalc)
library(MASS)


compile("fit.cpp")

dyn.load(dynlib("fit"))
#dyn.unload(dynlib("fit"))

compile("fit_sim.cpp")

dyn.load(dynlib("fit_sim"))
#dyn.unload(dynlib("fit_sim"))

compile("fit_dataparameters.cpp")

dyn.load(dynlib("fit_dataparameters"))
#dyn.unload(dynlib("fit_dataparameters"))

##############################################################################
l_grad = function(dataparameters,obj_dataparameters){
  
  return(-obj_dataparameters$gr(dataparameters))
}

##############################################################################

cAIC_calc = function(obj,opt,map,rname,sd.rep){
  
  p = length(opt$par)
  dati = obj$env$data
  
  par1 = obj$env$parList()
  parPriorMode <- obj$env$env$par
  parDataMode <- obj$env$last.par
  
  jnll = obj$env$f(parDataMode)
```

```
##Marginal Precision of random effects;
dati$wt = rep(0,nT)
obj_nd <- MakeADFun(dati, par1, map=map, DLL="fit",random=rname, silent=TRUE)
indx = obj$env$lrandom()
Hess = -Matrix(obj_nd$env$f(parDataMode,order=1,type="ADGrad"),sparse = TRUE)
cov_Psi_inv = -Hess[indx,indx]
q = nrow(cov_Psi_inv)

#-sum(dnorm(tmb.data$y, rep$Nay,rep$sd_err, TRUE))

ddlr.f = Hess[!indx,!indx]
ddlr.r = Hess[indx,indx]
ddlr.fr = Hess[!indx,indx]
ddlr.r_inv = -solve(cov_Psi_inv)
cov_Psi = -ddlr.r_inv;

## Joint hessian etc
hess.m = -Matrix(obj$env$f(parDataMode,order=1,type="ADGrad"),sparse = TRUE)
ddlj.f = hess.m[!indx,!indx]
ddlj.r = hess.m[indx,indx]
ddlj.rf = hess.m[indx,!indx]
ddlj.fr = t(ddlj.rf)

## conditional hessian and nll
ddlc.r = ddlj.r-ddlr.r
cnll = jnll - obj_nd$env$f(parDataMode)

## other calcs
ddlj.r_inv = as.matrix(solve(ddlj.r));
dPsi_dtheta = -ddlj.r_inv%*%ddlj.rf

cov_rtheta = sd.rep$cov.fixed
cov_rPsi =  -ddlj.r_inv + dPsi_dtheta%*%cov_rtheta%*%t(dPsi_dtheta)

domega_dtheta = rbind(diag(p),dPsi_dtheta)
t2 =  as.matrix(bdiag(matrix(0,p,p),-ddlj.r_inv) +
domega_dtheta%*%cov_rtheta%*%t(domega_dtheta))

cAIC = 2*cnll + 2*(p-1) + 2*q - 2*sum(diag((ddlj.r_inv)%*%ddlr.r))

## code for implementing Eq. (4)
rep_now = obj$report()

dati_dataparameters = dati
dati_dataparameters$y = NULL
par1_dataparameters = par1
par1_dataparameters$y = dati$y
data_vec = dati$y
names(data_vec) = rep("y",length(data_vec))
```

```r
  parDataMode_dataparameters = c(parDataMode,data_vec)
  ind_y = (names(parDataMode_dataparameters)=="y")
  
  dati_dataparameters$wt = rep(1,nT)
  obj_dataparameters <- MakeADFun(dati_dataparameters, par1_dataparameters, map=map,
                                  DLL="fit_dataparameters",random=rname, silent=TRUE)
  ind_RE = obj_dataparameters$env$lrandom()
  myenv <- new.env()
  myenv$dataparameters <- parDataMode_dataparameters[!ind_RE]
  myenv$obj_dataparameters <- obj_dataparameters
  nD <- numericDeriv(quote(l_grad(dataparameters,obj_dataparameters)),
c("dataparameters"), myenv)
  hess_l = attr(nD,"gradient")
  dtheta_dy = -hess_l[(p+1):(dim(hess_l)[1]),1:p]%*%ginv(hess_l[1:p,1:p])
  Hess_jointy =
-Matrix(obj_dataparameters$env$f(parDataMode_dataparameters,order=1,type="ADGrad"),sparse = TRUE)
  dRE_dy = -( Hess_jointy[ind_y,ind_RE]+dtheta_dy%*%ddlj.fr )%*%ddlj.r_inv
  dRE_dy = as.matrix(dRE_dy)
  dall_dy = cbind(dtheta_dy,dRE_dy)
  
  corr_dy =
2*exp(2*opt$par["log_sd_err"])*sum(diag(dall_dy%*%as.matrix(Hess_jointy[!ind_y,ind_y])))
  ##############################################################################
  
  ret = c(cAIC,corr_dy,2*(p-1) + 2*q - 2*sum(diag((ddlj.r_inv)%*%ddlr.r)))
  names(ret) = c("cAIC","Eq. (4)","Eq. (6)")
  
  return(ret)
}

################################################################################

ar1sim = function(n = 50, rho, u0 = 0, sd.e = 1, details = FALSE, seed = NULL){
  set.seed(seed)
  e = rnorm(n, 0, sd=sd.e) # random error
  u = numeric(n) # Generiere Nullvektor
  u[1] = u0 # Erste Beobachtung
  for (t in 2:(n)){ # Restliche Beobachtungen
    u[t] = rho * u[t-1] + e[t]
  }
  out = list(u.sim = u, n = n, rho = rho, e.sim = e)
  
  attr(out, "details") = if (details) {TRUE} else {FALSE}
  attr(out, "type") = "ar1sim"
  class(out) = c("desk")
  return(out)
```

```r
}
T = 50
age=1:6
A=max(age)
phi_year=0.8

## fixed age correlations in model m2;
phi.vec = c(0.01,0.1,0.4,0.8)
logit.phi.vec = log(phi.vec)
nphi = length(phi.vec)

nrep.vec = c(3,5)
tau.vec = c(1,2,4)

beta=0

niter=500

r.names = paste('iter=',1:niter,sep='')
m.names = c ("p","nll","mAIC","true BC","cAIC","Eq. (4)","Eq. (6)")
phi.names = c ("0.01", "0.1", "0.4", "0.8")
nrep.names = c("3","5")
tau.names = c("1","2","4")
sim.out = array(NA,c(length(nrep.vec),length(tau.vec),nphi,niter,7),
                dimnames=list (nrep.names, tau.names, phi.names, r.names, m.names))

for(nrepi in 1:length(nrep.vec)){
  for(taui in 1:length(tau.vec)){
    for(k in 1:length(phi.vec)){
      seed = 1
      error_count = 0
      while(seed <= niter){

        nrep = nrep.vec[nrepi]
        tau = tau.vec[taui]

        sigma_u = 1
        sigma_e = sigma_u/tau

        year.eff = ar1sim(n = T, rho = phi_year, u0 = beta, sd.e = sigma_u)$u.sim

        year.eff1=year.eff;

        age.eff = seq(-2,3,by=1)

        simdata=expand.grid(age=age,year=1:T,rep=1:nrep)

        simdata$year.eff = year.eff[simdata$year]
        simdata$age.eff = age.eff[simdata$age]
```

```r
simdata$mu =  simdata$year.eff + simdata$age.eff

nT = length(simdata$mu)
simdata$Y = simdata$mu + sigma_e*rnorm(nT,0)

tmb.data = list(y=simdata$Y,iy=simdata$year-1,ia=simdata$age-1,T=T)
tmb.data$wt = rep(1,nT)

rname = c("Ny")

parameters = list(
  log_sd_err=log(sigma_e),
  log_sd_N=log(sigma_u),
  logit_phi_year=log(phi_year/(1-phi_year)),
  log_sd_age=logit.phi.vec[k],
  q=age.eff,
  Ny=cbind(matrix(0,T,A),year.eff)
)


map = list(
)
map.m2=map

tmb.data$pool=0

## M2 Simulation ##
objm2 <- MakeADFun(tmb.data, parameters, map=map, DLL="fit",random=rname,
                   control = list(trace=0),silent = TRUE)

dat <- objm2$simulate();
tmb.data$y = dat$y
Nay_true = dat$Nay_true
#####################

tmb.data$pool=0
objm2 <- MakeADFun(tmb.data, parameters, map=map, DLL="fit",random=rname,
                   control = list(trace=0),silent = TRUE)

optm2<-nlminb(objm2$par,objm2$fn,objm2$gr,
              control = list(trace=0,eval.max=2000,iter.max=10000))

if (optm2$convergence != 0) {
  error_count = error_count + 1
  next
}

optm2$convergence = 100
```

```r
          sd.rep.m2 = sdreport(objm2)
          rep.m2 = objm2$report(objm2$env$last.par)


################################################################################
          tmb.data_sim = list(
            y=dat$y,
            iy=simdata$year-1,
            ia=simdata$age-1,
            T=T,
            sd_err_true = sigma_e,
            Nay_true = Nay_true,
            Nay_est = rep.m2$Nay,
            iter_inner = 1000
          )

          parameters_sim = list(
            log_sd_err=optm2$par["log_sd_err"],
            log_sd_N=optm2$par["log_sd_N"],
            logit_phi_year=optm2$par["logit_phi_year"],
            log_sd_age=optm2$par["log_sd_age"],
            q=optm2$par["q"],
            Ny=rep.m2$Ny
          )

          objm2_sim <- MakeADFun(tmb.data_sim, parameters_sim, map=map,
DLL="fit_sim",#random=rname,
                                 control = list(trace=0),silent = TRUE)

          dat_sim <- objm2_sim$simulate();


################################################################################

          sim.out[nrepi,taui,k,seed,1]=length(optm2$par)
          sim.out[nrepi,taui,k,seed,2]=optm2$objective
          sim.out[nrepi,taui,k,seed,3]=2*optm2$objective + 2*length(optm2$par)
          sim.out[nrepi,taui,k,seed,4]=-dat_sim$bias_corr
          sim.out[nrepi,taui,k,seed,5:7]=cAIC_calc(objm2,optm2,map.m2,rname,sd.rep.m2)

          seed = seed+1
      }
      print(error_count)
      save.image(file='sim.RData')
    }}
  print(nrepi)
}

tmp1=apply(sim.out,c(1,2,3,5),mean)
```

```r
library(reshape2)
l <- melt(tmp1)
colnames(l) = c("nrep","tau","phi","method","value")

tmp2 = dcast(l, nrep + tau + phi ~ method)

(tmp2["Eq. (6)"]-tmp2["true BC"])/tmp2["true BC"]
(tmp2["Eq. (4)"]-tmp2["true BC"])/tmp2["true BC"]

# save.image(file='sim.RData')
```

```cpp
//fit.cpp

#include <TMB.hpp>
template<class Type>
Type objective_function<Type>::operator() ()
{

  DATA_VECTOR(y);
  //   DATA_VECTOR(keep); //dataset.keep=keep;
  //   DATA_VECTOR_INDICATOR(keep, y); //dataset.keep=keep;
  DATA_IVECTOR(iy);
  DATA_IVECTOR(ia);
  DATA_INTEGER(T);
  DATA_VECTOR(wt);
  PARAMETER(log_sd_err);
  PARAMETER(log_sd_N);
  PARAMETER(logit_phi_year);
  PARAMETER(log_sd_age);
  PARAMETER_VECTOR(q);
  PARAMETER_MATRIX(Ny);
  DATA_INTEGER(pool);

  int n = y.size();
  int nage = q.size();
  Type one = 1.0;

  using namespace density;

  Type nll = Type(0.0);

  Type sd_err = exp(log_sd_err);
  Type sd_N = exp(log_sd_N);
  Type phi_year = exp(logit_phi_year)/(one + exp(logit_phi_year));
  Type sd_age = exp(log_sd_age);

  vector<Type> Nay(n);
  for(int i = 0;i < n;++i){
    if(pool==0){
      Nay(i) = sd_age*Ny(iy(i),ia(i)) + q(ia(i)) + sd_N*Ny(iy(i),nage);
    }
    if(pool==1){
      Nay(i) = q(ia(i)) + sd_N*Ny(iy(i),0);
    }
    nll -= wt(i)*dnorm(y(i), Nay(i), sd_err, true);
  }

  if(pool==0){
    array<Type> Ny_block(T,nage);
    for(int i = 0;i < nage;++i){
      Ny_block.col(i) = Ny.col(i);
```

```
    }
    nll += AR1(Type(0.0),AR1(Type(0.0)))(Ny_block);
    vector<Type> del = vector<Type>(Ny.col(nage));
    nll += AR1(phi_year)(del);
  }
  if(pool==1){
    vector<Type> del = vector<Type>(Ny.col(0));
    nll += AR1(phi_year)(del);
  }

  SIMULATE {
    vector<Type> Nay_true(n);
    array<Type> Ny_sim(T,nage);
    AR1(Type(0.0),AR1(Type(0.0))).simulate(Ny_sim);
    vector<Type> Ny_year(T);
    AR1(phi_year).simulate(Ny_year);
    for(int i = 0;i < n;++i){
      Nay_true(i) = sd_age*Ny_sim(iy(i),ia(i)) + q(ia(i)) + sd_N*Ny_year(iy(i));

      y(i) = rnorm(Nay_true(i), sd_err);  // Simulate Nay_true
    }
    REPORT(Nay_true);
    REPORT(y);
  }

  REPORT(q);
  REPORT(Ny);
  REPORT(sd_N);
  REPORT(sd_err);
  REPORT(sd_age);
  REPORT(phi_year);
  REPORT(Nay);
  ADREPORT(Ny);
  return nll;
}
```

```cpp
//fit_dataparameters.cpp

#include <TMB.hpp>
template<class Type>
Type objective_function<Type>::operator() ()
{

  //   DATA_VECTOR(y);
  //   DATA_VECTOR(keep); //dataset.keep=keep;
  //   DATA_VECTOR_INDICATOR(keep, y); //dataset.keep=keep;
  DATA_IVECTOR(iy);
  DATA_IVECTOR(ia);
  DATA_INTEGER(T);
  DATA_VECTOR(wt);
  PARAMETER(log_sd_err);
  PARAMETER(log_sd_N);
  PARAMETER(logit_phi_year);
  PARAMETER(log_sd_age);
  PARAMETER_VECTOR(q);
  PARAMETER_MATRIX(Ny);
  DATA_INTEGER(pool);
  PARAMETER_VECTOR(y);

  int n = y.size();
  int nage = q.size();
  Type one = 1.0;

  using namespace density;

  Type nll = Type(0.0);

  Type sd_err = exp(log_sd_err);
  Type sd_N = exp(log_sd_N);
  Type phi_year = exp(logit_phi_year)/(one + exp(logit_phi_year));
  Type sd_age = exp(log_sd_age);

  vector<Type> Nay(n);
  for(int i = 0;i < n;++i){
    if(pool==0){
      Nay(i) = sd_age*Ny(iy(i),ia(i)) + q(ia(i)) + sd_N*Ny(iy(i),nage);
    }
    if(pool==1){
      Nay(i) = q(ia(i)) + sd_N*Ny(iy(i),0);
    }
    nll -= wt(i)*dnorm(y(i), Nay(i), sd_err, true);
  }

  if(pool==0){
    array<Type> Ny_block(T,nage);
    for(int i = 0;i < nage;++i){
```

```
      Ny_block.col(i) = Ny.col(i);
    }
    nll += AR1(Type(0.0),AR1(Type(0.0)))(Ny_block);
    vector<Type> del = vector<Type>(Ny.col(nage));
    nll += AR1(phi_year)(del);
  }
  if(pool==1){
    vector<Type> del = vector<Type>(Ny.col(0));
    nll += AR1(phi_year)(del);
  }

  SIMULATE {
    array<Type> Ny_sim(T,nage);
    AR1(Type(0.0),AR1(Type(0.0))).simulate(Ny_sim);
    vector<Type> Ny_year(T);
    AR1(phi_year).simulate(Ny_year);
    for(int i = 0;i < n;++i){
      Nay(i) = sd_age*Ny_sim(iy(i),ia(i)) + q(ia(i)) + sd_N*Ny_year(iy(i));

      y(i) = rnorm(Nay(i), sd_err);   // Simulate Nay
    }
    REPORT(Nay);
    REPORT(y);
  }

  REPORT(q);
  REPORT(Ny);
  REPORT(sd_N);
  REPORT(sd_err);
  REPORT(sd_age);
  REPORT(phi_year);
  REPORT(Nay);
  ADREPORT(Ny);
  return nll;
}
```

```cpp
//fit_sim.cpp

#include <TMB.hpp>
template<class Type>
Type objective_function<Type>::operator() ()
{

  DATA_VECTOR(y);
  //  DATA_VECTOR(keep); //dataset.keep=keep;
  //  DATA_VECTOR_INDICATOR(keep, y); //dataset.keep=keep;
  DATA_IVECTOR(iy);
  DATA_IVECTOR(ia);
  DATA_INTEGER(T);
  //  DATA_VECTOR(wt);
  DATA_SCALAR(sd_err_true)
    DATA_VECTOR(Nay_true);
  DATA_VECTOR(Nay_est);
  DATA_INTEGER(iter_inner);

  PARAMETER(log_sd_err);
  PARAMETER(log_sd_N);
  PARAMETER(logit_phi_year);
  PARAMETER(log_sd_age);
  PARAMETER_VECTOR(q);
  PARAMETER_MATRIX(Ny);
  //  DATA_INTEGER(pool);

  int n = y.size();
  int nage = q.size();
  Type one = 1.0;

  using namespace density;

  Type nll = Type(0.0);

  Type sd_err = exp(log_sd_err);
  //Type sd_N = exp(log_sd_N);
  //Type phi_year = exp(logit_phi_year)/(one + exp(logit_phi_year));
  //Type sd_age = exp(log_sd_age);

  for(int i = 0;i < n;++i){
    nll -= dnorm(y(i), Nay_est(i), sd_err, true);
  }

  Type nll_star = Type(0.0);
  vector<Type> y_star(n);
  Type iter_inner_Type = 0;
  Type bias_corr;

  SIMULATE {
```

```
      for(int iter = 0;iter < iter_inner;++iter){
        iter_inner_Type += 1;
        for(int i = 0;i < n;++i){
          y_star(i) = rnorm(Nay_true(i), sd_err_true);   // Simulate Nay_true
          nll_star -= dnorm(y_star(i), Nay_est(i), sd_err, true);
        }
      }
      nll_star = nll_star/iter_inner_Type;
      bias_corr = 2*(nll-nll_star);
      REPORT(bias_corr);
    }

    //REPORT(q);
    //REPORT(Ny);
    //REPORT(sd_N);
    //REPORT(sd_err);
    //REPORT(sd_age);
    //REPORT(phi_year);
    //REPORT(Nay);
    //ADREPORT(Ny);
    return nll;
  }
```


	%\item[R-package for  MYNEW routine:] R-package ÒMYNEWÓ containing code to perform the diagnostic methods described in the article. The package also contains all datasets used as examples in the article. (GNU zipped tar file)
	
	%\item[HIV data set:] Data set used in the illustration of MYNEW method in Section~ 3.2. (.txt file)

%\end{description}

%We hope you've chosen to use BibTeX!\ If you have, please feel free to use the package natbib with any bibliography style you're comfortable with. The .bst file Chicago was used here, and agsm.bst has been included here for your convenience. 

\end{document}